\documentclass[journal=cmatex,manuscript=article,layout=twocolumn]{achemso} 
\usepackage[version=3]{mhchem} 
\usepackage{graphicx}
\usepackage{epsfig}
\usepackage{multirow}
\usepackage{dcolumn}
\usepackage{bm}
\usepackage{subfigure}
\usepackage{color}
\usepackage{amsmath}
\usepackage{amssymb}
\usepackage{amsthm}
\usepackage{todonotes}
\usepackage{rotating}
\mciteErrorOnUnknownfalse

\author{Neelam Yedukondalu}
\email{nykondalu@gmail.com}
\affiliation{Department of Analytical and Structural Chemistry, CSIR-Indian Institute of Chemical Technology, Tarnaka, Hyderabad 500007, India}
\alsoaffiliation{Department of Geosciences, Center for Materials by Design, and Institute for Advanced Computational Science, State University of New York, Stony Brook, New York 11794-2100, USA}
\alsoaffiliation{Joint Photon Sciences Institute, Stony Brook University, Stony Brook, New York 11790-2100, USA}

\author{Tribhuwan Pandey}
\affiliation{Department of Physics, University of Antwerp, Groenenborgerlaan 171, B-2020 Antwerp, Belgium}
\email{tribhuwan.pandey@uantwerpen.be}

\author{S Chand Rakesh Roshan}
\affiliation{Rajiv Gandhi University of Knowledge Technologies, Basar, Telangana-504107, India}
\alsoaffiliation{Department of Physics, National Institute of Technology-Warangal, Telangana, India} 

\date{\today}

\title[An \textsf{achemso} demo]
{Effect of Hydrostatic Pressure on Lone Pair Activity and Phonon Transport in Bi$_2$O$_2$S}
\begin{document}

\begin{abstract}
Dibismuth dioxychalcogenides, Bi$_2$O$_2$Ch (Ch = S, Se, Te) are emerging class of materials for next generation electronics and thermoelectrics with an ultrahigh carrier mobility and excellent air stability. Among these, Bi$_2$O$_2$S is fascinating because of stereochemically active 6$s^2$ lone pair of Bi$^{3+}$ cation, heterogeneous bonding and high mass contrast of constituent elements. In this work, we systematically investigate the effect of hydrostatic pressure and its implications on lattice dynamics and phonon transport properties of Bi$_2$O$_2$S by employing first principles calculations along with the Boltzmann transport theory. The ambient $Pnmn$ phase exhibits a low average lattice thermal conductivity ($\kappa_l$) of 1.71 W-m/K at 300 K. We also find that Bi$_2$O$_2$S undergoes a structural phase transition from low symmetry ($Pnmn$) to a high symmetry ($I4/mmm$) structure around 4 GPa due to the Bi$^{3+}$ cation centering. Upon compression the lone pair activity of Bi$^{3+}$ cation is suppressed, which increases $\kappa_l$ by nearly 3 times to 4.92 W-m/K at 5 GPa for $I4/mmm$ phase. The calculated phonon lifetimes and Gr\"uneisen parameters show that anharmonicity reduces with increasing pressure due to further suppression of lone pair, strengthening of intra and inter molecular interactions, which raises the average room temperature $\kappa_l$ to 12.82 W-m/K at 20 GPa. Overall, the present study provides a comprehensive understanding of hydrostatic pressure effects on stereochemical activity of the Bi$^{3+}$ cation lone pair and its consequences on phonon transport properties of Bi$_2$O$_2$S. 
\end{abstract}

\section{Introduction}
Chemical intuitions are central for the development of cutting edge technologies to exploit the renewable energy resources for thermal energy management applications. For example, the lone pair electrons are important for realizing low lattice thermal conductivity ($\kappa_l$) by inducing strong anharmonicity in the system~\cite{zeier2016thinking, skoug2011role, tan2016rationally,du2017impact,Seshadri2022}. Similarly, layered structures promote strong bonding heterogeneity (combination of rigid and fluctuating sub-lattices in a material) and has important implications to enhance complex interdependent thermoelectric properties, such as high power factor and low $\kappa_l$.\cite{Pal2018,pandey2020ultralow}  Recently, dibismuth dioxychalcogenides, Bi$_2$O$_2$Ch (Ch = S, Se, Te), and their 2D counterparts\cite{Yang2019,Li2021,Liang2021,Tippireddy2021,Bo2022} have gained tremendous research interest due to their potential applications in thermoelectrics,\cite{Luu2015,Wang2021,Tippireddy2021} ferroelectrics,\cite{Wu2017} and optoelectronics.\cite{Yang2022,Tippireddy2021,Li2021,Zhang2015,Guo2019} Among these, Bi$_2$O$_2$S is unique due to stereochemically active 6$s^2$ lone pair of Bi$^{3+}$ cation, and hence, it crystallizes in the primitive orthorhombic structure (SG: $Pnmn$, Z=2 formula units (f.u.) per primitive cell), while Bi$_2$O$_2$(Se/Te) crystallizes in the body centred tetragonal (SG : $I4/mmm$, Z = 1 f.u. per primitive cell) anti-ThCr$_2$Si$_2$-type structure. These crystals are made of alternating Bi$_2$O$_2$, and chalcogen (S/Se/Te) layers, which are held together by weak electrostatic forces\cite{Wei2019,Guo2021} in contrast to van der Waals (vdW) interactions that are typically observed in layered materials.\cite{meng2019thermal,Yedu2011,Yedu2022} Infra-red and Raman spectra\cite{Xu2019} and strain effects on bulk and monolayer Bi$_2$O$_2$Ch crystals are systematically investigated to explore their possible applications in nano-electronics.\cite{Cheng2018}

A high pressure X-ray diffraction study\cite{Bu2020} revealed that Bi$_2$O$_2$S undergoes a structural transition from $Pnnm$ $\rightarrow$ $I4/mmm$ at 6.4 GPa due to displacive nature of off-centred Bi$^{3+}$ cation and then 2D $\rightarrow$ 3D structure upon further compression above 13.2 GPa owing to disappearance of the 6$s^2$ lone pair. Bi$_2$O$_2$Se is one of the most stable Sillen-type compounds under pressure with no structural transition being observed until 30 GPa\cite{Pereira2018} while Bi$_2$O$_2$Te is found to be stable till 50 GPa.~\cite{Wang2020-1} The intrinsic ultralow $\kappa_l$ behavior\cite{Wang2018,Song2019,Song2020,Guo2021} and thermoelectric properties\cite{Wang2018,Song2020} of Bi$_2$O$_2$Ch are studied using first principles calculations and Boltzmann transport theory. Effect of hydrostatic pressure on stereochemically active lone pair and its implications on lattice dynamics and phonon transport of Bi$_2$O$_2$S are scarce in the literature.

Here, we systematically investigate the crystal and electronic structure, elastic properties, lattice dynamics, and phonon transport properties for the ambient ($Pnnm$) and high pressure ($I4/mmm$) phases of Bi$_2$O$_2$S. By analyzing pressure-dependent static enthalpy, lattice, bond parameters, and elastic constants, we show that Bi$_2$O$_2$S undergoes a structural phase transition from low symmetry ($Pnmn$) to a high symmetry ($I4/mmm$) structure around 4 GPa.  By analysis of electron localization function (ELF) and electronic structure, we further show that under pressure the lone pair electrons of Bi$^{3+}$ cation are suppressed.  This suppression of lone pair electrons reduces anharmonicity at high pressure and hence $\kappa_l$ is enhanced.

\begin{table*}[!ht]
\caption{Ground state structural properties such as lattice constants (a, b, c, in \r{A}) and volume (V in \r{A}$^3$) of ambient ($Pnmn$) and high pressure ($I4/mmm$) phases of Bi$_2$O$_2$S obtained using DFT-D3 method and compared with the X-ray diffraction measurements\cite{Bu2020}.}
\label{table1}
\begin{tabular}{ccccc} \hline
Phase & Parameter    &    This work  &   Expt.$^a$    &  Others  \\ \hline
$Pnmn$  &a     &    3.931      &   3.874    &    3.85$^b$ ,3.9 $^c$  ,3.89$^d$, ,3.837$^e$ ,3.972$^f$ \\
(0 GPa) &b     &    3.861      &   3.840    &    3.89$^b$  ,3.87 $^c$   ,3.87 $^d$,  3.848$^e$,3.884 $^f$\\
        &c     &    11.948     &   11.916   &  11.97$^b$  ,12.05 $^c$   ,11.99$^d$ ,11.94 $^e$ ,12.079 $^f$ \\
&V   &    181.34     &   177.264  &      -   \\

$I4/mmm$ &a     &   3.787   &   3.7705   &   -  \\
(5.6 GPa)&c     &  11.713    &   11.6902   &-    \\
&V   &   167.97    &   166.211   &  - \\ \hline
\end{tabular}
\\ $^a$Ref.\cite{Bu2020} $^b$Ref.\cite{Wu2017} $^c$ Ref. \cite{Cheng2018} $^d$Ref.\cite{Ma2018} $^e$ Ref.\cite{Xu2019} $^f$Ref.\cite{Hu2020}  
\end{table*}

\section{Computational details and Methodology}
\subsection{Structure relaxation and electronic structure}
The first calculations have been performed using Vienna Ab-initio Simulation Package (VASP)\cite{Kresse1996}. The electron-electron interactions $i.e.$ exchange-correlation are treated by Generalized Gradient Approximation (GGA) within the parametrization of Perdew-Burke-Ernzerhof (PBE). The electron-ion interactions are treated with pseudopotential approach. We have used DFT-D3 method\cite{Grimme2011} to capture the weak electrostatic interactions. The following plane wave basis orbitals were considered as valence electrons; Bi: 5d$^{10}$,6s$^2$,6p$^3$; O: 2s$^2$,2p$^4$; and S: 3s$^2$,3p$^4$. A kinetic energy cutoff of 600 eV was used for plane wave basis set expansion and a spacing of 2$\pi$ $\times$ 0.025 \r{AA}$^{-1}$ for k-mesh in the irreducible Brillouin zone. The plane wave cutoff energy as well as k-spacing were varied first to ensure the total energy convergence. Self consistency criteria is chosen in such way that the total energy is converged to 1e$^{-8}$ eV/atom and the maximal force on each atom is less than 1e$^{-3}$ eV/\r{AA}. Electronic structure is calculated using Tran Blaha modified Becke Johnson (TB-mBJ) potential\cite{Tran2009} as implemented in VASP including spin-orbit coupling. Crystal structures and electron localization functions are visualized using the VESTA software\cite{VESTA2008}. Local Orbital Basis Suite Towards Electronic Structure Reconstruction (LOBSTER) package~\cite{Lobster2020} is used to perform crystal orbital Hamiltonian population (COHP) analysis.

\subsection{Interatomic force constants}
Harmonic and anharmonic interatomic force constants (IFCs) are needed to calculate phonon thermal conductivity $\kappa_l$ from first principles calculations. For both the phases, harmonic IFCs were obtained using the finite displacement method within the Phonopy package~\cite{togo2015first} with 5$\times$5$\times$2 (500 atoms) supercell and $\Gamma$-point only k-grid. Other parameters are same as those used in the structural relaxation. Long range corrections to phonon frequencies were included by calculation of Born effective charges and dielectric constants within the method proposed by Wang et. al.~\cite{wang2010mixed}. Anharmonic IFCs were calculated using 4$\times$4$\times$2 supercell. For the displaced supercells configurations phonon-phonon interactions were considered up to the 5.5 \r{AA} distance. This requires 900 and 500 DFT calculations for $Pnmn$ and $I4/mmm$ phases, respectively. The rest of the parameters are same as those used in the harmonic IFCs calculation, except a slightly smaller energy cutoff (520 eV) was used to reduce the computational cost. 

\subsection{Lattice thermal conductivity}
The calculated harmonic and anharmonic force constants were used to solve the phonon Boltzmann transport equation iteratively as implemented in the ShengBTE code~\cite{li2012thermal1,li2012thermal,li2014shengbte}. The lattice thermal conductivity due to particle like transport channel ($\kappa_p$) can be give as:
\begin{equation}
\label{eq:SCPH}
\kappa_p^{\alpha \beta} = \frac{1}{N\Omega}\sum_{\lambda} C_{\lambda}v_{\lambda, \alpha} \otimes v_{\lambda, \beta}\tau_{\lambda}
\end{equation}
where N, and $\Omega$ are the number of unitcells in the system and volume of the unitcell, respectively. $C_{\lambda}$,  $\tau_{\lambda}$ are the specific heat and phonon lifetimes for a phonon mode $\lambda$ with wave-vector \textbf{q} and  polarization j, respectively. $v_{\lambda, \alpha}$ is the $\alpha^{th}$ component of the phonon group velocity $v_{\lambda}$.

Convergence of $\kappa_l$ with respect to various dependent parameters such as cut-off distance for 3$^{rd}$ order IFCs, Gaussian smearing width was carefully tested and the same are provided in the supporting information (Figure S1). For the calculations reported here, a Gaussian smearing width of 1.0 was used to approximate the Dirac distribution, and a q-point mesh of 21$\times$21$\times$9 was used for Brillouin zone integration. In these calculations, three-phonon scattering and isotopic disorder scattering~\cite{tamura1983isotope} from natural isotope mass variation are considered. 

The solution of Boltzmann transport equation discussed above only includes the contribution from the particle transport channel. As recently shown for strongly anharmonic materials contribution from the coherence channel can also be significant~\cite{simoncelli2019unified,luo2020vibrational,xia2020microscopic, jain2020multichannel,pandey2022origin}. The coherence contribution ($\kappa_c$) to lattice thermal conductivity is calculated using the off diagonal terms of velocity operator~\cite{hardy1963energy,allen1993thermal} using the method proposed by Simoncelli and co-authors~\cite{simoncelli2019unified}. The $\kappa_c$ is described as
\begin{equation}
\begin{split}
\scriptsize
\kappa^{\alpha \beta}_{c} & =\frac{\hbar^2}{k_{B} {T}^2}\frac{1}{\Omega N_q}\sum_{\mathbf{q}}\sum_{j\neq j'}\frac{\omega_{\mathbf{q}j} +\omega_{\mathbf{q}{j'}}}{2} {V}^{j,j'}_{\mathbf{q},\alpha}.{V}^{j,j'}_{\mathbf{q},\beta}\\
& \frac{\omega_{\mathbf{q}j}n_{\mathbf{q}j}(n_{\mathbf{q}j} + 1) + \omega_{\mathbf{q}j'}n_{\mathbf{q}j'}(n_{\mathbf{q}j'} + 1)}{4(\omega_{\mathbf{q}j'}- \omega_{\mathbf{q}j})^2 + (\Gamma_{\mathbf{q}j} + \Gamma_{\mathbf{q}j'})^2}(\Gamma_{\mathbf{q}j} + \Gamma_{\mathbf{q}j'})
\end{split}
\end{equation}
where, $k_{B}$, $\hbar$, $\Omega$, $N_q$, and $n_{\mathbf{q}j}$ are the Boltzmann constant, reduced Plank constant, volume of the unit cell, number of $\mathbf{q}$ points,  the Bose-Einstein distribution, respectively. $\Gamma_{\mathbf{q}j}$ is the phonon linewidth and ${V}^{j,j'}_{\mathbf{q},\alpha}$ is the $\alpha^{th}$ component of velocity operator. The thermal conductivity results presented in the main article include contribution from both particle $\kappa_p$, and coherence $\kappa_c$ channels ($\kappa_l = \kappa_p + \kappa_c$). As shown in Figure S2 and S3, the contribution at room temperature, the $\kappa_c$ varies from 0.3 W/m-K (0 GPa) to 0.16 W/m-K (20 GPa) at high pressures.

\section{Results and Discussion}

\begin{figure}[!ht]
    \centering
    \includegraphics[width=\columnwidth]{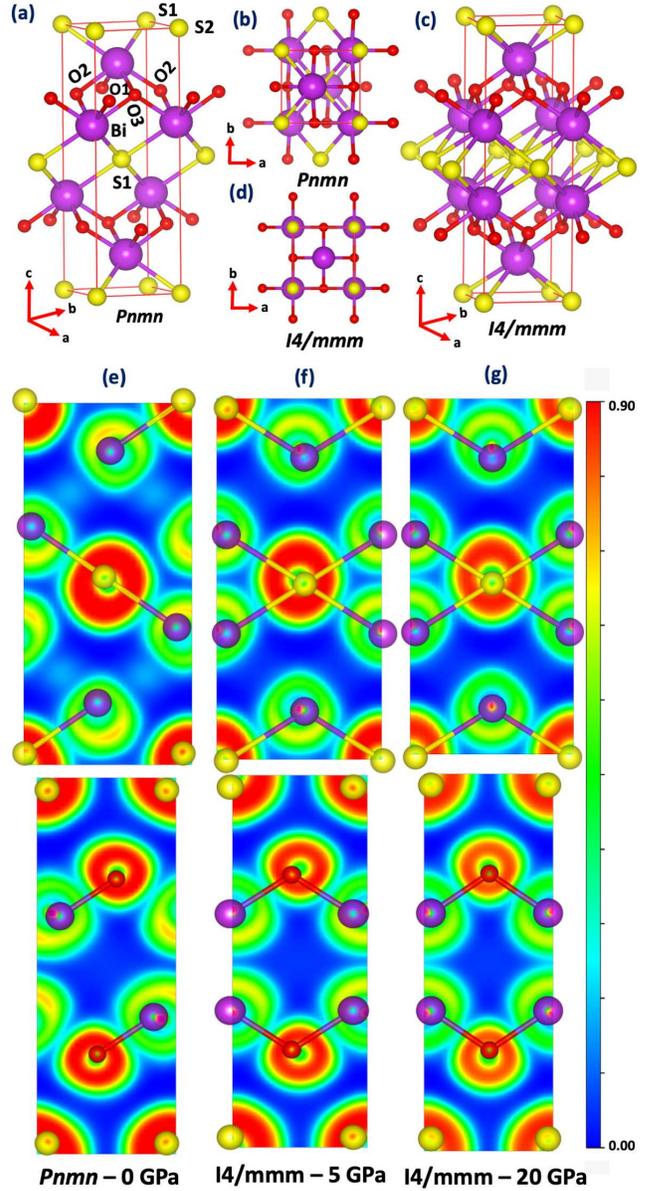}
    \caption{Crystal structure of (a) ambient ($Pnmn$) phase (b) distorted crystal structure in $ab$-plane due to lone pair at Bi$^{3+}$ cation (c) crystal structure of high pressure ($I4/mmm$) phase and (d) ordered crystal structure in $ab$-plane due to centering of Bi$^{3+}$ cation in $I4/mmm$ phase. Calculated electron localization function (ELF) at (e) 0 GPa for $Pnmn$ phase and (f) 5 and (g) 20 GPa for $I4/mmm$ phase projected on the (110) plane for Bi-S bonds (middle panel) and (010) plane for Bi-O bonds (bottom panel). Red and blue ends represent high electron localization and no localization, respectively. All ELFs are plotted at an iso-surface value of 0.005 e/\r{AA}$^3$.}
    \label{fig:str}
\end{figure}{}

\subsection{Crystal structure and structural phase transition}
As discussed in the introduction, Bi$_2$O$_2$S crystallizes in the low symmetry ($Pnmn$) structure (see Figures \ref{fig:str}(a-b)) compared to heavy chalcogen based systems namely Bi$_2$O$_2$Se and Bi$_2$O$_2$Te at ambient conditions. This is due to the presence of stereochemically active $6s^2$ lone pair of Bi$^{3+}$ cation in Bi$_2$O$_2$S while its activity might be suppressed by chemical pre-compression of heavy chalcogens (Se and Te) in Bi$_2$O$_2$Se and Bi$_2$O$_2$Te systems, which drives them to crystallize in relatively higher symmetry ($I4/mmm$) structure (see Figures \ref{fig:str}(c-d)) and the $Pnmn$ structure is distorted analogue of $I4/mmm$ crystal structure. To investigate the bonding differences between these two phases further, we examined the ELF~\cite{becke1990simple, seshadri2001visualizing} as depicted in Figures~\ref{fig:str}(e)-(g). The ELF is projected onto the (110) and (010) planes, which contain Bi, S and Bi, O atoms, respectively. The red color represents complete electron localization, while the blue color represents almost no electron localization. 

The ELF of the two phases appear similar at first glance, but significant differences can be seen around Bi-S and Bi-O bonds. In the $Pnmn$ phase, the Bi$^{3+}$ cation lone pair can be seen by yellow lobe regions of electron localization in Figure~\ref{fig:str}(e). When the pressure is applied (Figures ~\ref{fig:str}(f)-(g)), the intensity of the yellow lobes surrounding the Bi$^{3+}$ cation is significantly reduced, indicating that the lone pair of Bi$^{3+}$ cation is being suppressed. Due to presence of lone pair in $Pnmn$ phase, the coordination environment around Bi$^{3+}$ cation differs greatly between these two phases. In $Pnmn$ phase, Bi and S have two in-equivalent bond lengths and these are listed as Bi-S1 (3.00 \r{AA}) and Bi-S2 (3.51 \r{AA}) in Table S1 and Figure~\ref{fig:str}(a). While in the $I4/mmm$ phase, Bi and S bonds have four-fold coordination with only one in-equivalent bond with a length of 3.14 \r{AA}. Similarly, the bond length between Bi and O atoms in the $Pnmn$ phase ranges from 2.24 to 2.43 \r{AA} (Bi-O1, Bi-O2 and Bi-O3 as shown in Table S1 and Figure~\ref{fig:str}(a)), whereas in the $I4/mmm$ phase, all Bi-O bonds are identical, with a bond length of 2.19 \r{AA}. Overall, the bonding is highly heterogeneous in the $Pnmn$ phase over $I4/mmm$ phase. 

\begin{figure*}[!ht]
    \centering
   \includegraphics[width=0.95\textwidth]{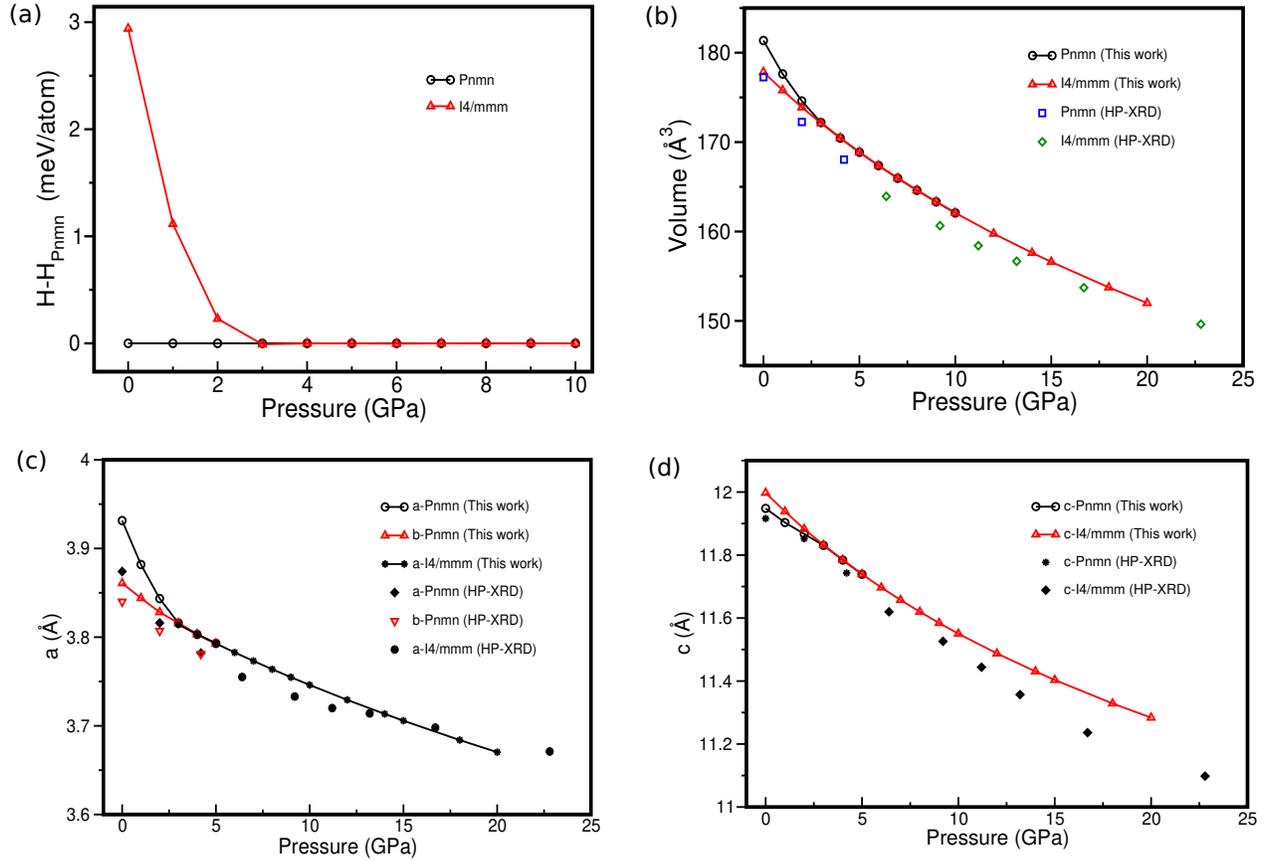}
    \caption{(a) Relative enthalpy difference of high pressure ($I4/mmm$) phase w.r.t ambient ($Pnmn$) phase. Pressure dependent (b) volume (V) (c, d) lattice constants (a, c) of ambient ($Pnmn$) and high pressure ($I4/mmm$) phases of Bi$_2$O$_2$S are compared with the lattice constants and volume determined from the HP-XRD measurements.\cite{Bu2020}}
    \label{fig:lattice}
\end{figure*}{}

A recent HP-XRD study\cite{Bu2020} revealed that Bi$_2$O$_2$S undergoes a structural phase transition from orthorhombic ($Pnmn$) to tetragonal ($I4/mmm$) structure at 6.4 GPa due to dynamic cation centering and then it transforms from 2D to 3D structure upon further compression above 13.2 GPa because of complete suppression of the $6s^2$ lone pair of Bi$^{3+}$ cation. To get further insights on structural phase transition and anharmonicity, we have systematically investigated the high pressure behavior of Bi$_2$O$_2$S up to 20 GPa. As shown in Figure~\ref{fig:lattice}(a), the enthalpy difference between ambient and high pressure phases is gradually decreasing with pressure up to 3 GPa and there after both phases exhibit iso-enthalpic nature above 4 GPa. This result suggests a 2$^{nd}$ order phase transition from $Pnmn$ $\rightarrow$ $I4/mmm$ in Bi$_2$O$_2$S. To gain further insights on the structural phase transition, we analyze the pressure dependent lattice constants, bond parameters as a function of pressure and equation of state (EOS). As shown in Figure \ref{fig:lattice}(b), the calculated volumes for both $Pnmn$ and $I4/mmm$ phases are converging and become almost equal above 3 GPa until the maximum studied pressure range (0-20 GPa) in this work and it is in excellent agreement with the HP-XRD measurements.\cite{Bu2020} No volume reduction is observed at the transition pressure which corroborates with iso-enthalpic nature of ambient and high pressure phases above 4 GPa. The calculated lattice constants decrease monotonically with increasing pressure. A slightly distinct variation of lattice constants above 10 GPa is observed from the HP-XRD measurements.\cite{Bu2020} Interestingly, the lattice constants \textit{a} and \textit{b} merge into a single lattice constant above 4 GPa and lattice constant \textit{c} merges into \textit{c} lattice constant for high pressure tetragonal ($I4/mmm$) phase (see Figure \ref{fig:lattice}(c,d)). This observation is consistent with the variation of lattice constants as a function of pressure as observed in the HP-XRD measurements.\cite{Bu2020} The calculated pressure coefficients by fitting the pressure versus lattice parameter data to a quadratic expression show that the $Pnmn$ lattice is highly compressible along a-crystallographic direction, which is consistent with the obtained low elastic moduli along the x-axis. 

As illustrated in Figure S4 and S5, the calculated in-equivalent bond lengths (Bi-S and Bi-O) and angles (S-Bi-S and O-Bi-O) of lower symmetry ambient $Pnmn$ phase converge and merge into a single in-equivalent bond length/angle for higher symmetry $I4/mmm$ phase above 4 GPa. From HP-XRD measurements,\cite{Bu2020} above 13.2 GPa, a sharp reduction of Bi-S and increase of Bi-O bonds have been observed due complete suppression of lone pair thus resulting in $I4/mmm$ structure to be transformed from 2D to 3D. However, such a sharp reduction in Bi-S and increase in Bi-O bond lengths are not observed using static first principles calculations under high pressure, instead these bond lengths are monotonically decreasing with pressure (Figure S4). This indicates that the lone pair of Bi$^{3+}$ cation was not completely suppressed until 20 GPa, This result is consistent with a combined study\cite{Bu2020} of HP-XRD and first principles calculations, where a relatively small activity of the lone pair for $I4/mmm$ phase could be seen at 22.8 GPa although lone pair was completely suppressed at 13.2 GPa from HP-XRD measurements of the same study. The possible reasons for the distinct behavior of static first principles calculations and HP-XRD measurements might be 1) HP-XRD measurements were performed using Neon as pressure transmitting medium,\cite{Bu2020} which might create a quasi-hydrostatic or non-hydrostatic conditions above 13 GPa.\cite{Klotz2009} 2) Anharmonic effects are not considered for computation of structural properties under high pressure whereas HP-XRD measurements were carried out at 300 K.\cite{Bu2020} The static enthalpy calculations and pressure evolution of structural properties of ambient and high pressure phases clearly demonstrate that Bi$_2$O$_2$S undergoes a continuous phase transition from $Pnmn$ $\rightarrow$ $I4/mmm$ under high pressure due to displacive nature of the off centred Bi$^{+3}$ cation (Figure S6) which removes the lattice distortion and probing high pressure phase to crystallizes in high symmetry structure under hydrostatic compression.

\begin{figure*}[!ht]
    \centering
   \includegraphics[width=0.95\textwidth]{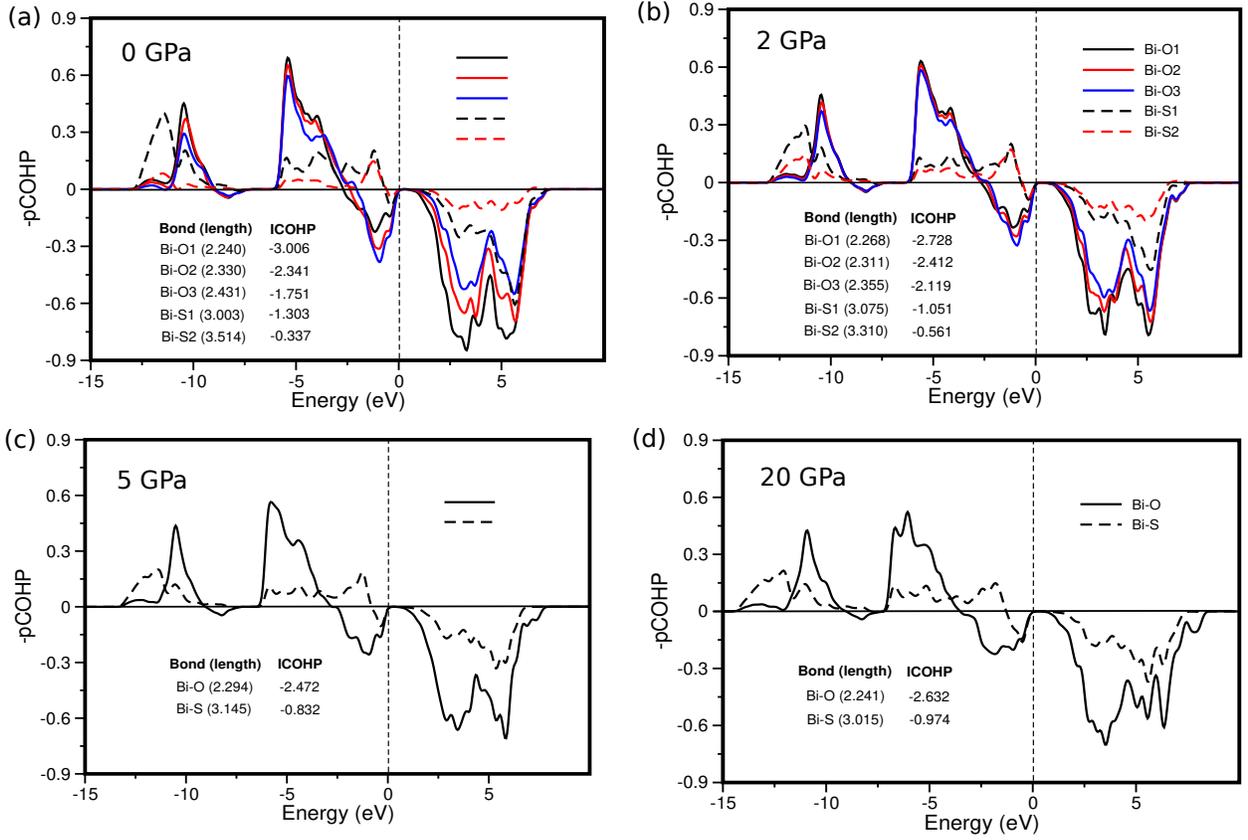}
    \caption{Projected crystal orbital Hamiltonian population (pCOHP) at (a) 0 GPa and (b) 2 GPa for $Pnmn$ phase and at (c) 5 GPa and (d) 20 GPa for $I4/mmm$ phase. The integrated COHP values are given at each pressure for both $Pnmn$ and $I4/mmm$ phases which show the variation of strength of the bonds as a function pressure as shown in Figure S7. The bonding states are positive and anti-bonding states are negative. Bond lengths are given in \r{AA} and ICOHP values provided in eV}
    \label{fig:pcohp}
\end{figure*}

To gain further insights into bonding changes under pressure we then analyze the projected CHOP (pCOHP) for various Bi-O and Bi-S bonds for both ambient and high pressure phases.  Figures \ref{fig:pcohp} (a)-(d) shows the pCHOP for in-equivalent bonds of $Pnmn$ and $I4/mmm$ phases as a function of pressure. There are significant anti-bonding states for Bi-O bonds compared to Bi-S bonds above and near the Fermi level. Relatively, the anti-bonding states in $I4/mmm$ phase are stabilizing with pressure compared to the ambient $Pnmn$ phase. The integrated COHP (ICOHP) analysis show that the $Pnmn$ phase consists of mixed strong (more negative ICOHP value indicates stronger bonding) and weak (less ICOHP negative value indicates weaker bonding) bonds of Bi-O and Bi-S at ambient pressure (Figure S7). Especially, Bi-S2 shows a relatively weaker bonding nature resulting in rattling behavior as shown from the atomic mean square displacements, which favours strong anahrmonicity thus lowering the $\kappa_l$.

\begin{table*}[!ht]
\caption{Calculated enhanced Born effective charges (Z*) of ambient ($Pnmn$) and high pressure ($I4/mmm$) phases of Bi$_2$O$_2$S.}
\label{Born}
\begin{tabular}{cccccccccc} \hline
 Atom    & Formal charge  &                   Z*($Pnmn$)                                      &    Z*($I4/mmm$)    \\ \hline
 Bi      &    +3          &   ${\begin{pmatrix}  5.34 & 0.00 & 0.19  \\ 0.00  &  5.89 &  0.00 \\  0.54 & 0.00 &  5.19  \\ \end{pmatrix}}$   &   ${\begin{pmatrix}  5.83 & 0.00 & 0.00  \\ 0.00  &  5.83 & 0.00 \\ 0.00 & 0.00 &  5.24  \\ \end{pmatrix}}$ \\
 O       &    -2          &  ${\begin{pmatrix} -3.55 & 0.00 & 0.58  \\ 0.00  & -3.73 &  0.00 \\  0.10 & 0.00 & -3.61 \\ \end{pmatrix}}$     & ${\begin{pmatrix} -3.69 & 0.00 & 0.00  \\ 
 0.00  & -3.69 &  0.00 \\ 0.00 & 0.00 & -3.71 \\  \end{pmatrix}} $ \\
 S       &    -2          &  ${\begin{pmatrix} -3.57 & 0.00 & 0.37  \\  0.00  & -4.31 &  0.00 \\ 0.94 & 0.00 & -3.15 \\ \end{pmatrix}}$     & ${\begin{pmatrix} -4.28 & 0.00 & 0.37  \\
0.00  & -4.28 &  0.00 \\ 0.94 & 0.00 & -3.05 \\ \end{pmatrix}}$ \\   \hline
\end{tabular}
\end{table*}

\subsection{Electronic structure and Born effective charges}
It is well-known fact that the standard DFT functionals underestimate the band gaps\cite{Heyd2005,Verma2017} by 30-40 $\%$ for semiconductors and insulators due to self-interaction error (SIE), to overcome SIE problem, we have used TB-mBJ\cite{Tran2009} potential to calculate electronic structure of Bi$_2$O$_2$S. The calculated band gaps are 0.78 and 0.60 eV without and with inclusion of spin-orbit coupling (SOC) within PBE-GGA, respectively for $Pnmn$ phase and the corresponding band gaps obtained using TB-mBJ potential are 1.55 and 1.45 eV. The obtained PBE-GGA band gap 0.78 eV without SOC is comparable with previously obtained band gap of 1.0 eV,\cite{Zhang2015} and the TB-mBJ band gap with SOC 1.45 eV is consistent with HSE06 band gap of 1.25 eV.\cite{Wu2017} As shown by Shi et al.~\cite{Shi2016}, materials with fully occupied lone pair ($ns^2$, n = 4, 5, 6) show distinct features in their electronic structure by having an additional band (dominated by fully occupied cation $s$-states) below the valence band compared to the materials without lone pair cation. The conduction band is mainly dominated by extended $p$-states of lone pair cation in contrast to the cation $s$-states in materials without lone pair cation. The calculated projected electronic density of states (PDOS) for both ambient orthorhombic and high pressure tetragonal (at 5 GPa, 10 GPa and 20 GPa) phases of Bi$_2$O$_2$S are presented in Figure S8. The band profiles look similar for ambient and high pressure phases at 0 GPa and 5 GPa, respectively. As discussed above, the bottom of the valence band is mainly derived from 6$s$-states of Bi$^{3+}$ cation for both of these phases ($Pnmn$ and $I4/mmm$). As shown in Figure S8, the top of the valence band is mainly dominated by the $p$-states of S$^{2-}$ and O$^{2-}$ anions and also minor contribution from $s$ and $p$-states of the Bi$^{3+}$ cation, while the conduction band is mainly dominated by extended $p$-states of the Bi$^{3+}$ cation, which hybridize with anion (S$^{2-}$ and O$^{2-}$) $p$-states of the valence band, thus, results in a mixed ionic-covalent character which causes a significant cross-band-gap hybridization in Bi$_2$O$_2$S. The observed mixed ionic-covalent character causes strong lattice polarization for the lone pair cation containing compounds.\cite{Du2010,Du2010InX} This could be clearly seen from Born effective charges (BECs), as presented in Table \ref{Born}, the calculated BECs are significantly enhanced for both phases of Bi$_2$O$_2$S compared to the formal charges of cation (+3) and anions (-2). Enhanced BECs provide large LO-TO splitting from phonon dispersion curves which brings the lattice (TO modes) in the proximity of ferroelectric instability\cite{Du2010} thereby lowering $\kappa_l$~\cite{pandey2020lattice}. Therefore, it is very intriguing to investigate elastic and phonon transport properties of stereochemically active lone pair materials under hydrostatic compression in general and in particular for Bi$_2$O$_2$S. 

\subsection{Mechanical and dynamical stability}
To explore the mechanical and dynamical stability, we calculated second order elastic constants (SOECs) and phonon dispersion curves, respectively for $Pnmn$ and $I4/mmm$ phases of Bi$_2$O$_2$S at ambient as well as at high pressure. The calculated SOECs at 0 and 5 GPa for both the phases are given in Table \ref{elastic}. At 0 GPa, the $Pnmn$ phase obeys necessary and sufficient conditions of Born stability criteria\cite{Mouhat2014} and no imaginary frequencies are observed, which indicate that the $Pnmn$ phase is mechanically and dynamically stable at ambient pressure (see Figure \ref{fig:PHDISP}(a)). The $I4/mmm$ phase doesn't satisfy the Born stability criteria (C$_{44}$ $>$ 0) and also it possesses imaginary frequencies along $\Gamma$-M direction (see Figure \ref{fig:PHDISP}(b)) at 0 GPa, which clearly demonstrate that $I4/mmm$ phase is mechanically and dynamically unstable at ambient pressure which clearly demonstrates that role stereochemical activity of the lone pair in determining the stability of $Pnmn$ phase at ambient pressure for Bi$_2$O$_2$S.

\begin{table*}[!ht]
\caption{Calculated second order elastic constants (in GPa) of ambient ($Pnmn$) and high pressure ($I4/mmm$) phases of Bi$_2$O$_2$S at 0 and 5 GPa.}
\label{elastic}
\begin{tabular}{ccccccccccc} \hline
Phase  & Pressure &  C$_{11}$  &  C$_{22}$  &  C$_{33}$   &  C$_{44}$  &   C$_{55}$   &  C$_{66}$   &  C$_{12}$  & C$_{13}$  & C$_{23}$  \\ \hline
$Pnmn$ &  0.0    &  57.8      &   126.0    &  119.5      &   16.6     &   27.8       &   22.1      &   24.7     &    23.4   &  28.3     \\
$I4/mmm$ & 0.0   &  160.5     &   160.5    &  139.4      &   -6.79    &   -6.79      &   60.5      &   74.0     &    41.0   &  41.0     \\ 
$Pnmn$ & 5.0     &  189.4     &   189.0    &  184.9      &   14.7     &   14.7       &   72.6      &   96.6     &    57.1   &  57.3     \\
$I4/mmm$ & 5.0   &  188.3     &   188.3    &  184.0      &   14.7    &   14.7        &   72.6      &   95.8     &    57.8   &  57.8     \\ \hline
\end{tabular}
\end{table*}

When a non-zero hydrostatic pressure is applied to a particular crystal system, the Born stability criteria must be revised with modified elastic constants at a given hydrostatic pressure (P) conditions. According to Sinko and Smirnov\cite{Sinko2002,Sinko2004} the modified elastic constants at a given pressure are $\widetilde{C}_{ii}$ = $C_{ii}$ - P, (for $i$ = 1-6) and $\widetilde{C}_{12}$ = $C_{12}$ + P,  $\widetilde{C}_{13}$ = $C_{13}$ + P and $\widetilde{C}_{23}$ = $C_{23}$ + P. Therefore, the modified Born stability criteria under hydrostatic pressure for orthorhombic crystal symmetry can be found in our previous work,\cite{Yedu2015} whereas for tetragonal system the same are given as follows:

\begin{equation}
\begin{gathered}
 C_{11} - C_{12} - 2P  >  0;  C_{44} - P  >  0\\
P^2 + P (C_{11} + C_{12} + 4C_{13})\\ + 2C^2_{13} - C_{33}(C_{11} + C_{12}) > 0;   
C_{66} - P > 0      
\end{gathered}
\end{equation}

Under high pressure ($>$ 4 GPa), for instance at 5 GPa (see Table \ref{elastic}), the calculated SOECs for both $Pnmn$ and $I4/mmm$ phases have the following relationships: C$^{Pnmn}_{11}$ $\approx$ C$^{Pnmn}_{22}$ $\approx$ C$^{I4/mmm}_{11}$, C$^{Pnmn}_{44}$ = C$^{Pnmn}_{55}$ = C$^{I4/mmm}_{44}$ and C$^{Pnmn}_{12}$ $\approx$ C$^{Pnmn}_{13}$ = C$^{I4/mmm}_{12}$. These relationships strongly suggest a continuous phase transition from $Pnmn$ to $I4/mmm$ is occurring due to displacive nature of Bi$^{3+}$ cation in Bi$_2$O$_2$S with increasing pressure (Figure S6) and is in good accord with the pressure dependent structural properties such as pressure dependent lattice constants and bond parameters (Figures \ref{fig:lattice}, S3 and S4). In addition, no imaginary frequencies are found for the computed phonon dispersion curves at 5 GPa, which strongly suggest that $I4/mmm$ phase becomes dynamically stable above 5 GPa (Figure S9).   

\subsection{Lattice dynamics and phonon transport}

\begin{figure*}[!ht]
    \centering
    \includegraphics[width=0.8\textwidth]{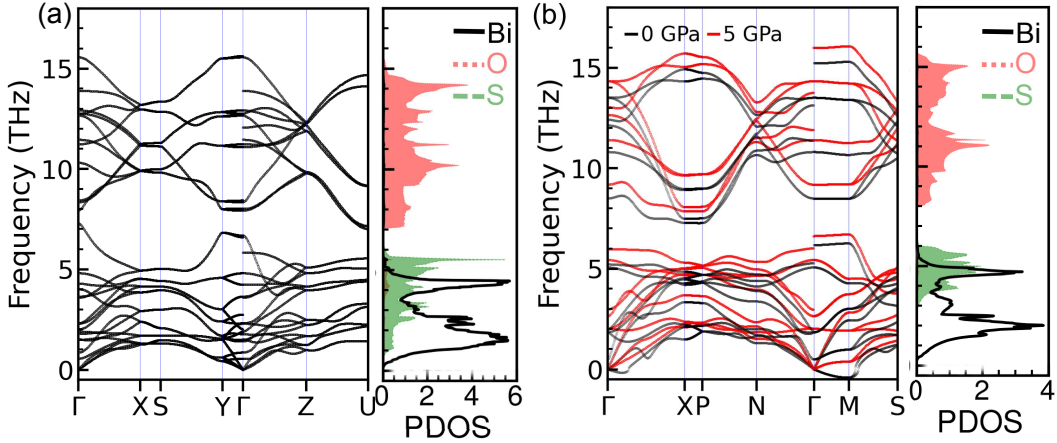}
    \caption{Phonon dispersion curves and density of states for Bi$_2$O$_2$S in (a) ambient ($Pnmn$) and (b) high pressure ($I4/mmm$) phases. For $I4/mmm$ phase, the 0 GPa phonon dispersion is also shown. As can be seen under applied pressure, $I4/mmm$ phase becomes dynamically stable.}
    \label{fig:PHDISP}
\end{figure*}{}

In general, the lone pair electrons are shown to favor low lattice thermal conductivity~\cite{nielsen2013lone,du2017impact}, and the pressure tunability of lone pair activity can have important implications for $\kappa_l$. To investigate this further, we analyze the lattice dynamics and phonon transport of Bi$_2$O$_2$S as a function of pressure within the 0-20 GPa pressure range. Note that below 5 GPa pressure, Bi$_2$O$_2$S has orthorhombic $Pnmn$  structure and at higher pressure it transforms to tetragonal $I4/mmm$ structure. Therefore, at low pressure (0, and 2 GPa) $\kappa_l$ calculations are performed for $Pnmn$ structure, whereas at high pressure ($\geq$ 5 GPa) $\kappa_l$ calculations are done for $I4/mmm$ structure. The calculated phonon dispersion curves and phonon density of states (PHDOS) are shown in Figures~\ref{fig:PHDISP}(a) and (b) for 0 GPa ($Pnmn$) and 5 GPa ($I4/mmm$) phases, respectively. Phonon dispersion for other pressure values are shown in the supporting information Figure S9. For both $Pnmn$ and $I4/mmm$ phases, the three acoustic modes are comprised of very low frequencies ($\leq$ 1.5 THz) and they overlap with optical branches. In both the phases, the optical phonons show very low frequency modes due to presence of heavy metal Bi and they hybridize with the acoustic phonon modes. The optical phonon branches in high-frequency regions ($>$ 6 THz) are dispersive and exhibit large phonon group velocity, indicating the possibility of contribution to $\kappa_l$. The PHDOS in these compounds can be divided into three main regions: Bi dominated region below 1 THz, mixed Bi and S region above 1 THz, and O dominated region above 6 THz. Despite the huge mass contrast between Bi and S (m$_{Bi}$/m$_S$ = 6.5), there is a significant overlap between Bi and S in the PHDOS. This overlap is larger for the $Pnmn$ phase (1 THz - 5.5 THz) than in the $I4/mmm$ phase (3 THz - 5.5 THz). The sulfur PHDOS in the $Pnmn$ phase shows several small peaks in 1 THz - 5.5 THz and a large peak around 5 THz. The $I4/mmm$ phase also shows sulfur peaks in the PHDOS, but these are less pronounced. The large overlap between acoustic and optical phonons can lead to enhanced scattering of the heat-carrying acoustic phonons --- suggesting the possibility of relatively low $\kappa_l$ for the $Pnmn$ phase. 

\begin{figure*}[!ht]
    \centering
    \includegraphics[width=0.9\textwidth]{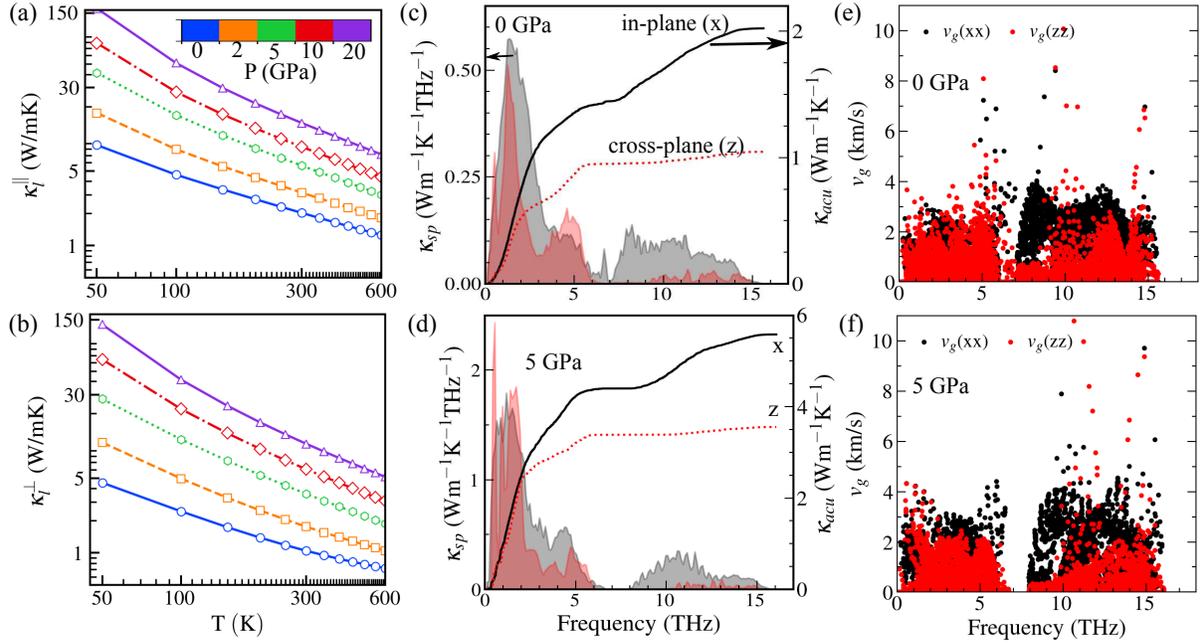}
    \caption{Comparison of calculated $\kappa_l$ as a function of temperature for ambient and 2 GPa $Pnmn$ and high pressure 5, 10 and 20 GPa $I4/mmm$ phases of Bi$_2$O$_2$S along (a) in-plane ($\parallel$) and cross-plane ($\perp$) directions. Calculated room temperature spectral thermal conductivity $\kappa_{sp}$ (shaded areas) and thermal conductivity accumulation $\kappa_{acu}$ (curves) as a function of phonon frequencies at (c) 0 GPa and (d) 5 GPa.  In plane (xx) and cross plane (zz) components of phonon group velocities as a function of frequencies at (e) 0 GPa and (f) 5 GPa.}
    \label{fig:Kappa}
\end{figure*}{}

The temperature-dependent $\kappa_l$ at various pressures is compared in Figure~\ref{fig:Kappa}(a) and (b) along in-plane and cross-plane directions. As can be seen there, $\kappa_l$ of the ambient  $Pnmn$ phase is smaller than the high pressure $I4/mmm$ phase. At room temperature, the average $\kappa_l$ values are 1.71, 4.92 W/m-K  and 12.65 W/m-K for $Pnmn$ (0 GPa), $I4/mmm$ (5 GPa), and $I4/mmm$ (20 GPa) phase, respectively. The calculated 300 K lattice thermal conductivity values ($\kappa_l^{\parallel}$= 2.02 W/m-K; $\kappa_l^{\perp}$ = 1.046 mW/m-K) for the ambient phase are somewhat smaller than the ones reported ( $\kappa_l^{\parallel}$ = 2.93 W/m-K; $\kappa_l^{\perp}$ = 1.7 W/m-K) in a previous study~\cite{zhang2019significant}. This difference largely stems from different lattice parameters (due to different in DFT functional), and different cutoffs used for calculations of 3$^{rd}$ order IFCs. At 2 GPa hydrostatic pressure the in-plane lattice parameters are similar to the ones reported in reference~\cite{zhang2019significant}, and the calculated lattice thermal conductivity ($\kappa_l^{\parallel}$ = 3.13 W/m-K, and $\kappa_l^{\perp}$ = 1.75 W/m-K) is in better agreement with previously reported values.~\cite{zhang2019significant} When compared with the $\kappa_l$ of with other dibismuth dioxychalcogenides such as Bi$_2$O$_2$Se~\cite{Guo2021} ($\kappa_l^{\parallel}$ = 1.71 W/m-K; $\kappa_l^{\perp}$ = 0.81 W/m-K at 300 K) the calculated room temperature $\kappa_l$ for the ambient Bi$_2$O$_2$S are a bit higher. This difference in $\kappa_l$ originates from the lower group velocity induced by relatively heavier Se atom in Bi$_2$O$_2$Se. Nonetheless the  $\kappa_{\perp}$ of ambient Bi$_2$O$_2$S are comparable with well know bismuth chalcogenides thermoelectric materials such as, Bi$_2$Te$_3$~\cite{witting2019thermoelectric}, Bi$_2$Se$_3$~\cite{paulatto2020thermal,Guo2021}, and Bi$_2$S$_3$~\cite{biswas2012tellurium,pandey2016simultaneous}.

Since both $Pnmn$ and $I4/mmm$ phases have an anisotropic crystal structure, the same is reflected in the calculated $\kappa_l$. Due to the orthorhombic structure, the $Pnmn$ phase exhibits slightly different $\kappa_l$ along in-plane \textit{x} (2 W/m-K) and \textit{y} (2.12 W/m-K) directions. Upon application of pressure as the $a$ and $b$ lattice parameters converge to single value, $\kappa_l$ along in-plane x and y direction also becomes same (Figure S2 (b)). Upon further application of pressure, the pressure $\ge$ 5 GPa, $\kappa_l$ becomes identical along x and y directions (Figure S2 (c)). As expected for layered materials with weak electrostatic interactions,\cite{Guo2021} the $\kappa_l$ along the cross-plane direction is smaller than the in-plane direction. Interestingly, the calculated thermal transport anisotropy is much smaller than what is typically observed in vdW layered materials such as, MoS$_2$ where $\kappa^{\parallel}_l$/$\kappa^{\perp}_l$  of nearly 30 was found~\cite{meng2019thermal}. This clearly distinguishes difference between the role of interlayer weak electrostatic and vdW bonding in determining the thermal anisotropy within the layered materials.  

\begin{figure}[!ht]
    \centering
    \includegraphics[width=0.9\columnwidth]{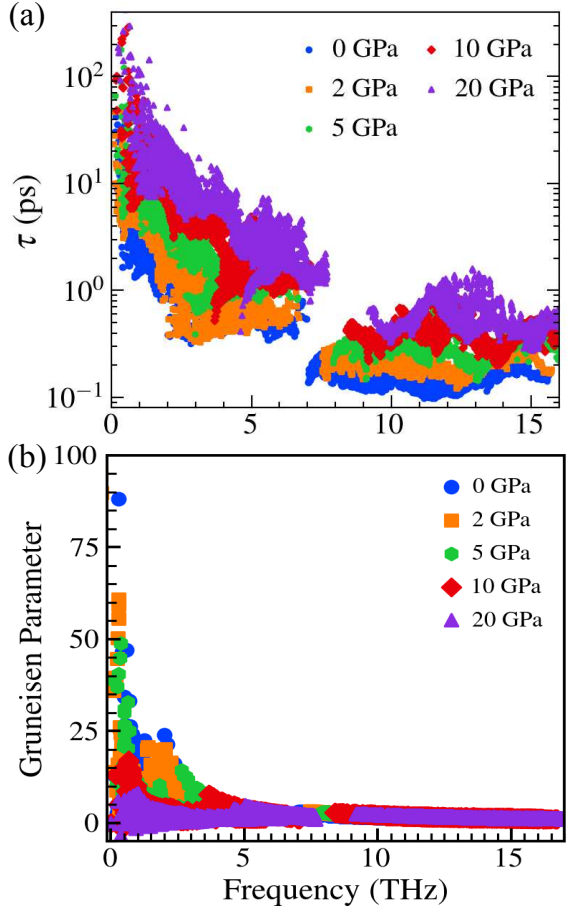}
    \caption{(a) Comparison of room temperature phonon lifetime ($\tau$) for ambient (0 and 2 GPa, $Pnmn$) and high pressure (5, 10 and 20 GPa $I4/mmm$) phase as a function of phonon frequencies. (b) Variation of room temperature Gr\"uneisen parameter as a function of phonon frequencies at selected pressures.}
    \label{fig:Kappa2}
\end{figure}

To understand the thermal transport behavior further we calculate the cumulative $\kappa_{acu}$ and spectral $\kappa_{sp}$ lattice thermal conductivity as a function of phonon frequency, which is shown in Figure~\ref{fig:Kappa}(c) and (d) for $Pnmn$ and $I4/mmm$ phases.  As can be seen there, in both the phases, the optical phonons contribute significantly to $\kappa_l$ particularly along the in-plane direction. The phonon frequency range contributing to cross-plane and in-plane $\kappa_l$ is also quite different. The $\kappa_l^{\perp}$ is mostly originates from phonons with frequency $<$ 5 THz, whereas phonons with frequencies up to 12 THz contribute to $\kappa_l^{\parallel}$ in both the phases. This can be understood by analyzing phonon group velocities which are shown in the Figure~\ref{fig:Kappa2}(e) and (f) for $Pnmn$ and $I4/mmm$ phases, respectively. In both the phases for the majority of the frequency range $v_g$ is higher along in-plane (xx) direction than the cross-plane (zz) direction. This is more prominent for optical phonons in the frequency window of 6-12 THz, where in-plane $v_g$ is more than two times larger than the cross-plane $v_g$. These higher in-plane $v_g$ values are the main reason for higher $\kappa_l$ along the in-plane direction. This is true for all the pressures investigated here as shown in Figure S10.

{}

We then focus on exploring the origin of the relatively low $\kappa_l$ in the $Pnmn$ phase. To explain the calculated differences in $\kappa_l$, we look into three factors, phonon group velocity, specific heat, and three phonon scattering rates, that determine $\kappa_l$. As explained above, the phonon group velocities (Fig.~\ref{fig:Kappa} (e) and (f)) are comparable in both phases --- therefore phonon velocities are not the reason for the lower $\kappa_l$ in the $Pnmn$ phase. As per the second-factor specific heat, the $Pnmn$ phase has only slightly higher ($\sim$ 1.3 \%) specific heat than the $I4/mmm$ phase. This indicates that the difference in phonon anharmonicity should be the main reason for different $\kappa_l$. The phonon anharmonicity governs the strength of phonon-phonon interactions which determines the overall intrinsic thermal resistance. Greater anharmonicity results in shorter phonon lifetimes and lower $\kappa_l$. Indeed, as shown in Figure~\ref{fig:Kappa2}(a), the phonon lifetimes of the $Pnmn$ phase are significantly smaller than the 5 GPa $I4/mmm$ phase over the entire frequency. The optical phonons in $Pnmn$ phase are more dispersive than in $I4/mmm$ phase, which enables it to satisfy the conservation laws required for three phonon scattering resulting in higher scattering phase space and therefore shorter phonon lifetimes. With increasing pressure the phonon lifetimes continue increasing, resulting in enhancement $\kappa_l$ with pressure. This is also consistent with the calculated Gr\"uneisen parameters which also decrease with increasing pressure (Figure~~\ref{fig:Kappa2}(b)) --- signifying reduction in anharmonicity. 

Both $\kappa_l^{\parallel}$ and $\kappa_l^{\perp}$ monotonically increase with applied pressure, and at 20 GPa, the room temperature $\kappa_l$ becomes --- $\kappa_l^{\parallel}$ = 14 W/m-K and $\kappa_l^{\perp}$ = 10.5 W/m-K. This increase of $\kappa_l$ with pressure is mainly driven by enhancement of phonon lifetimes under pressure (Fig.~\ref{fig:Kappa2} (a)) as the enhancement on both in-plane and cross-plane phonon group velocity due to applied pressure is rather small (Figure S10). $\kappa_l$ anisotropy ($\kappa_l^{\parallel}/\kappa_l^{\perp}$) also reduces with applied pressure from 1.94 (0 GPa) to 1.3 (20 GPa) at 300 K (Figure S11). This reduction in $\kappa_l$ anisotropy indicates enhanced inter layer interactions, which is in good agreement with the recent HP-XRD experiments.\cite{Bu2020}

\begin{figure*}[!t]
    \centering
    \includegraphics[width=0.85\textwidth]{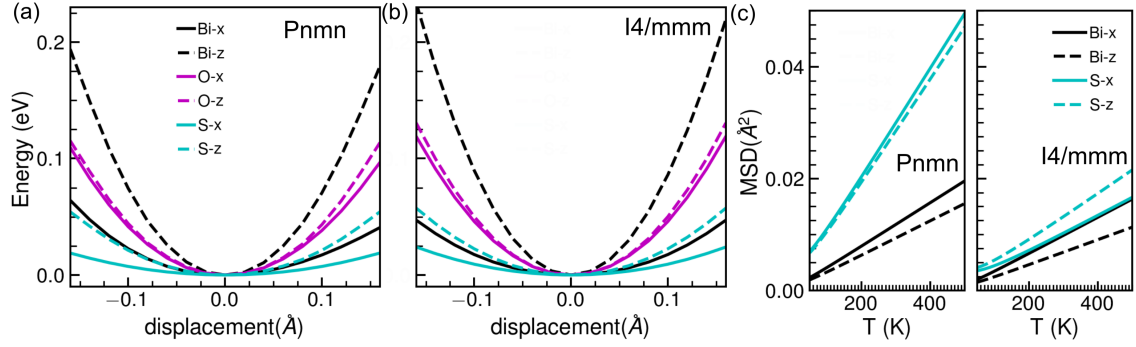}
    \caption{Average potential energy with respect to displacements of Bi, O and S atoms along Cartesian x and y directions in (a) $Pnmn$ and (b) $I4/mmm$ phases of Bi$_2$O$_2$S. (c) calculated average mean square displacements along Cartesian x and y directions for Bi and S atoms in both $Pnmn$ and $I4/mmm$ phase as a function of temperature. As can be seen mean square displacements are significantly larger for $Pnmn$ phase.}
    \label{fig:PES}
\end{figure*}{}

As discussed earlier, the main difference between orthorhombic $Pnmn$ and tetragonal $I4/mmm$ phases is the suppression of Bi$^{3+}$ 6$s^2$ lone pair electrons in the $I4/mmm$ phase, which may explain its relatively less anharmonicity.  To gain further insights into the role of lone pair electrons in bonding and anharmonicity, we next calculate the potential energy by off-centering the Bi, O, and S atoms along Cartesian x and z directions. The resulting potential energy plots are presented in Figures~\ref{fig:PES}(a) and (b) for $Pnmn$ and $I4/mmm$ phases, respectively. As can be seen there, the displacement potential energy of Bi and S atoms are significantly different along x and z directions in both $Pnmn$ and $I4/mmm$ phases. We find that for Bi and S atomic displacements in the x or y directions, the potential energy curve is relatively flat, whereas displacement along z-direction requires more energy. In both phases the energy required for O displacement is comparable. 

The potential energy analysis stipulates that Bi and S vibrational modes could be easily excited by low energy. This is consistent with the calculated PHDOS, and PHDOS of Bi and S atoms show significant overlap despite huge mass difference. Furthermore, for Bi vibration in $Pnmn$ phase potential energy curve shows asymmetric features and deviates from parabolic (harmonic) behavior. The polynomial fit of this potential energy well gives anharmonic coefficient as 2.99 and 1.68 along x and z directions, respectively. Such large deviation from typical parabolic behavior corroborates the strong anharmonicity of this phase. For comparison,  in $I4/mmm$ phase the anharmonic coefficients for Bi atomic vibrations along z direction is 1.56, and for Bi displacement along x direction the potential is harmonic (coefficient of anharmonic fit is zero). The above displacement potential energy analysis confirms that the $Pnmn$ phase is more anharmonic than $I4/mmm$ phase.  This is further confirmed by plotting the mean square displacements (MSD) for Bi and S atoms which are shown in Figure~\ref{fig:PES}(c) for both $Pnmn$ and $I4/mmm$ phases. The MSD for Bi and S atoms are higher in $Pnmn$ phase  (Bi = 0.012 \r{AA}$^2$, and S = 0.024 \r{AA}$^2$ at 300 K), than in the $I4/mmm$ phase (Bi = 0.009 \r{AA}$^2$, and S = 0.011 \r{AA}$^2$ at 300 K). This is expected because the longer Bi-S2 bond lengths in $Pnmn$ phase make the atoms move easily from their equilibrium positions during the vibration than in the $I4/mmm$ phase. Overall, the anharmonicity of the $6s^2$ lone pair of Bi$^{3+}$ cation plays a pivotal role in determining the phonon transport in Bi$_2$O$_2$S and it is suppressed (in $I4/mmm$ phase) upon compression as can be seen from chemical pre-compression of Se/Te atoms in Bi$_2$O$_2$Se and Bi$_2$O$_2$Te.\cite{Wei2019,Guo2021}

\section{Conclusions}
In summary, we have systematically investigated the effect of hydrostatic pressure on stereochemically active 6$s^2$ lone pair containing Bi$_2$O$_2$S with weak electrostatic interlayer bonding, and their implications on crystal structure, lattice dynamics and phonon transport properties using first principles calculations and Boltzmann transport theory. We predict that Bi$_2$O$_2$S undergoes a continuous structural phase transition from low symmetry distorted ($Pnmn$) structure to a high symmetry ordered ($I4/mmm$) structure at around 4 GPa, which is comparable to the transition pressure 6.4 GPa determined from the HP-XRD measurements\cite{Bu2020}. The mechanical and dynamical stability of both $Pnmn$ and $I4/mmm$ phases are confirmed from the calculated elastic constants and phonon dispersion curves, respectively. The obtained enhanced Born effective charges (BECs) clearly demonstrate that both the phases show significant cross-band-gap hybridization from their electronic structure suggesting near ferroelectric instability and is favorable to achieve low $\kappa_l$. The obtained low average $\kappa_l$ value 1.71 W-m/K for $Pnmn$ phase over 4.93 W-m/K for $I4/mmm$ phase is due to suppression of lone pair of Bi$^{3+}$ cation at high pressure which reduces anharmonicity in $I4/mmm$ phase. Both $\kappa_l^{\parallel}$ and $\kappa_l^{\perp}$ monotonically increase with applied pressure and become 14 W/m-K, and 10.5 W/m-K at 20 GPa, respectively. The highly anharmonic behavior of Bi and S atoms of $Pnmn$ phase demonstrated from PES and MSD at ambient pressure is mainly responsible for low $\kappa_l$. Overall, the present work provides an insight on how hydrostatic pressure effects the chemical bonding, lattice dynamics and phonon transport in Bi$_2$O$_2$S. The present study could stimulate exploring pressure dependent phonon transport properties in materials especially with lone pair cation, which will provide a new avenue of pressure modulated low $\kappa_l$ materials for future thermal energy applications.

\section{Acknowledgments}
NYK and TP contributed equally to this manuscript. NYK would like to thank Institute for Advanced Computational Science, Stony Brook University for providing computational resources (Seawulf cluster). TP is supported by Research Foundation Flanders (FWO-Vl). SCRR would like to thank RGUKT Basar for providing computational facilities.

\section{Supporting Information}
Convergence tests of $\kappa_l$ w.r.t cutoff distance and Gaussian broadening (Fig. S1), coherence contribution to $\kappa_l$ (Fig. S2), Phonon life time and mean free paths (Fig. S3), pressure dependent variation of bond parameters (Figs. S4 and S5), centering of Bi$^{3+}$ cation with pressure (Fig. S6), pressure dependent variation of ICOHP (Fig. S7), Electronic and phonon band structures of $Pnmn$ and $I4/mmm$ phases at high pressure phases (Figs. S8 and S9),   phonon group velocities at high pressure (Fig. S10) and pressure variation of thermal transport anisotropy (Fig. S11) (PDF).

\bibliography{Refs.bib}


\begin{figure*}
    \centering
    \includegraphics[width=6.5in,height=4.0in]{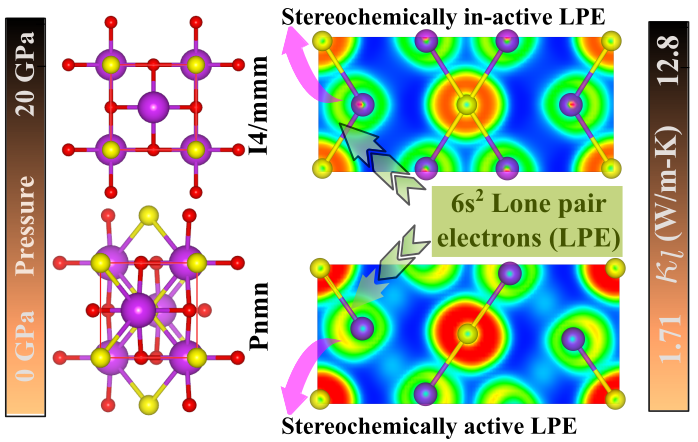}
    \caption*{Effect of hydrostatic pressure on stereochemically active 6$s^2$ lone pair of Bi$^{3+}$ cation in Bi$_2$O$_2$S ($Pnmn$) is systematically investigated. Bi$_2$O$_2$S transformed to Bi$_2$O$_2$Se/Te-type ($I4/mmm$) structure due to dynamic centering of Bi$^{3+}$ cation. The active and in-active nature of lone pair has profound implications on phonon transport, which aid for in silico design of advanced energy conversion materials.}
    \label{fig:TOC}
\end{figure*}

\end{document}


\clearpage

\begin{figure*}
    \centering
    \includegraphics[width=0.8\columnwidth]{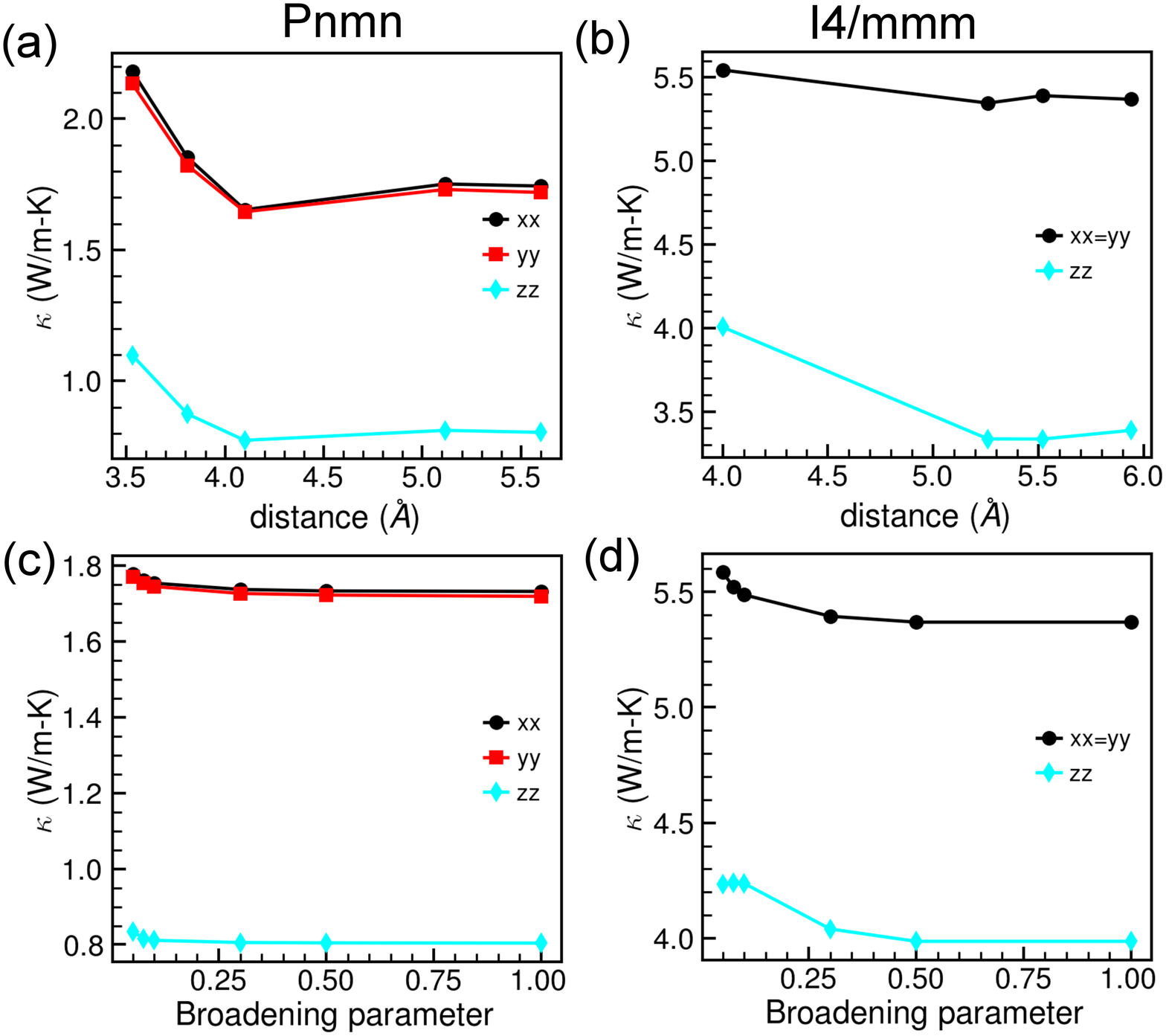}
   \caption{Convergence of calculated lattice thermal conductivity (without coherence term) with respect to cutoff distance for 3$^{rd}$ order interatomic force constants for (a) 0 GPa $Pnmn$ (b) 5 GPa $I4/mmm$ phases of Bi$_2$O$_2$S.  Convergence of calculated lattice thermal conductivity with respect to Gaussian broadening parameters at (a) 0 GPa for $Pnmn$ phase (b) 5 GPa $I4mmm$ phases of Bi$_2$O$_2$S. These calculations are performed at k-mesh of 21$\times$21$\times$7, and for panel (a) and (b) broadening parameter of 1.0 was used.}
\label{fig:BL}
\end{figure*}{}

\begin{figure*}
    \centering
    \includegraphics[width=0.8\columnwidth]{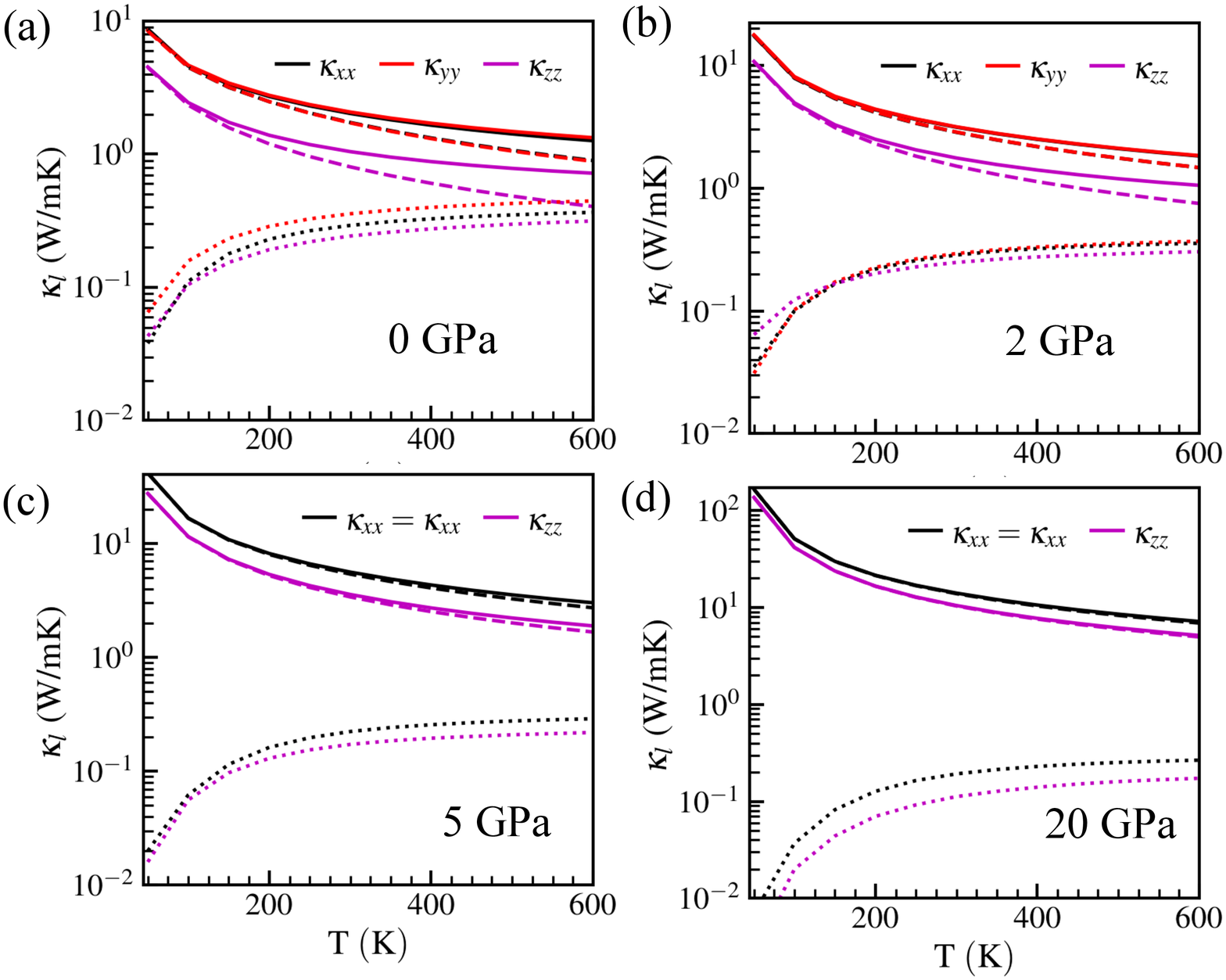}
    \caption{Calculated temperature dependent thermal conductivities of  Bi$_2$O$_2$S at (a) 0 GPa, (b) 2 GPa, (c) 5 GPa and (d) 20 GPa pressures. Here dashed, and dotted lines represent thermal conductivity due to particle ($\kappa_p$) and coherence ($\kappa_c$) channels, respectively.  Summation of $\kappa_l$ and $\kappa_c$ gives total lattice thermal conductivity from both channels.}
    \label{fig:kappa_c}
\end{figure*}

\section{Coherence thermal conductivity}
At 0 GPa,  room  temperature $\kappa_c$ value is $\sim$ 0.3 W/m-K, and the same is reduced to 0.17 W/m-K at 20 GPa. The calculated $\kappa_c$ contributions can be understood by analysis of phonon mean free path ($\Lambda$) and phonon lifetimes ($\tau$) as a function of phonon frequency which are shown in Figure S3. Within the Ioffe Regal limit, phonons are characterized as well-defined in space (time) when mean free paths are greater than the average atomic spacing a ($\Lambda > a$) (lifetimes are greater than the reciprocal angular frequency $\tau > 1/\omega$)~\cite{simoncelli2019unified,luo2020vibrational,pandey2022origin}.  These phonons mainly contribute to particle like thermal conduction ($\kappa_p$). Whereas phonons with $\Lambda < a$ and $\tau < 1/\omega$ contribute to coherence conductivity ($\kappa_c$). As shown below is Figure S3 (a), at all pressure values phonon are well-defined in time ( $\tau > 1/\omega$) and only few phonons modes satisfy $\Lambda < a$ (Figure S3 (b)). When the pressure is applied $\Lambda$ further increases and even fewer modes satisfy $\Lambda < a$. Therefore, $\kappa_c$ contribution decreases with applied pressure.

\begin{figure*}
    \centering
    \includegraphics[width=0.8\columnwidth]{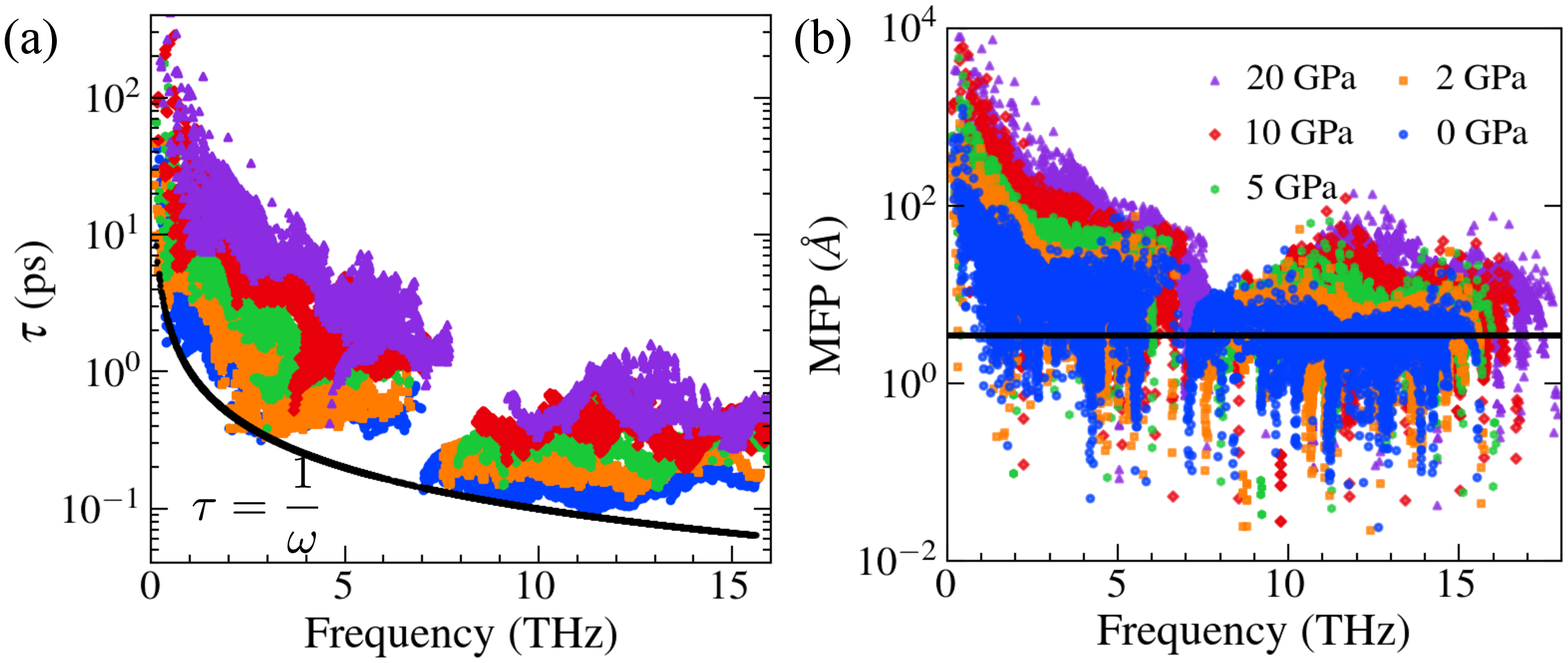}
    \caption{Room temperature (a) phonon lifetime ($\tau$) and (b) phonon mean free paths ($\Lambda$) as a function of phonon frequency. Black curve in (a)  give the phonon lifetimes in the Ioffe Regel limit ($\tau=1/\omega$). Black black horizontal line is the average bond length “a”.}
    \label{fig:kappa_c}
\end{figure*}

\begin{figure*}
    \centering
    \includegraphics[width=0.8\columnwidth]{Figures/FigS1.eps}
   \caption{Calculated in-equivalent bond lengths of ambient ($Pnmn$) and high pressure ($I4/mmm$) phases of Bi$_2$O$_2$S as a function of pressure.}
    \label{fig:BL}
\end{figure*}{}

\begin{figure*}
    \centering
    \includegraphics[width=0.6\columnwidth]{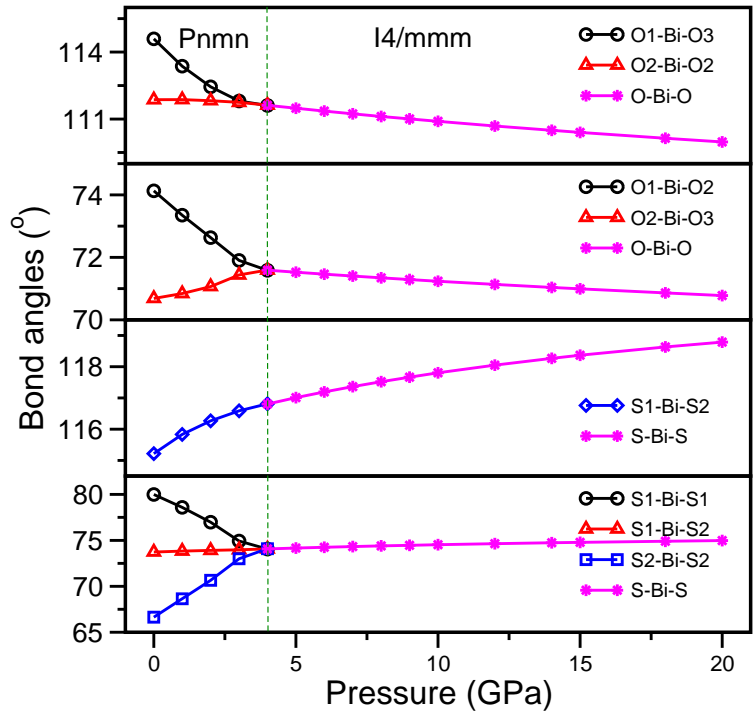}
   \caption{Calculated in-equivalent bond angles of ambient ($Pnmn$) and high pressure ($I4/mmm$) phases of Bi$_2$O$_2$S as a function of pressure.}
    \label{fig:BL}
\end{figure*}{}

\begin{figure*}
    \centering
    \includegraphics[width=0.6\columnwidth]{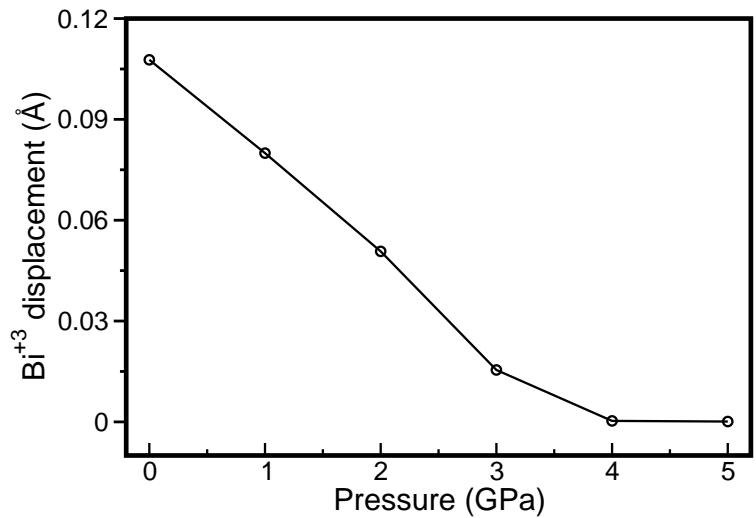}
   \caption{Centering of Bi$^{3+}$ cation in ambient ($Pnmn$) phase of Bi$_2$O$_2$S as a function of pressure.}
    \label{fig:Bi}
\end{figure*}{}

\begin{figure*}
    \centering
    \includegraphics[width=0.6\columnwidth]{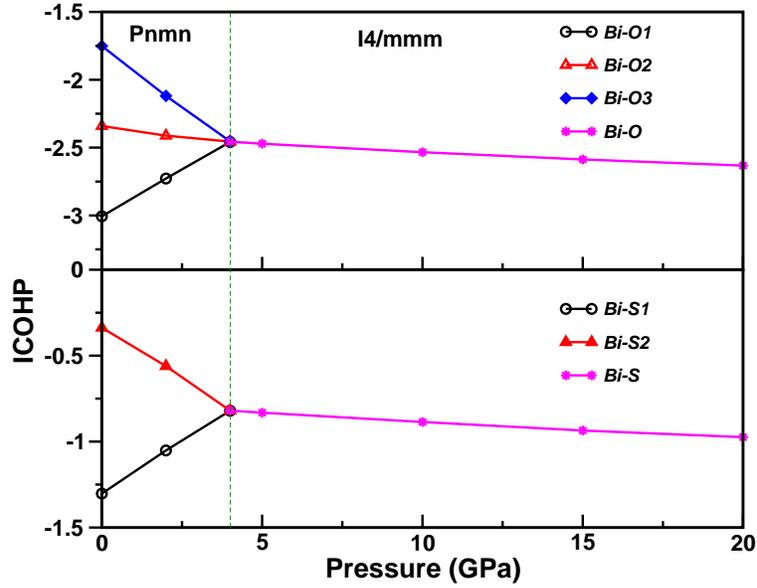}
   \caption{Calculated ICOHP values for in-equivalent bond lengths of ambient ($Pnmn$) and high pressure ($I4/mmm$) phases of Bi$_2$O$_2$S as a function of pressure.}
    \label{fig:ICOHP}
\end{figure*}{}

\begin{figure*}
    \centering
    \includegraphics[width=0.95\columnwidth]{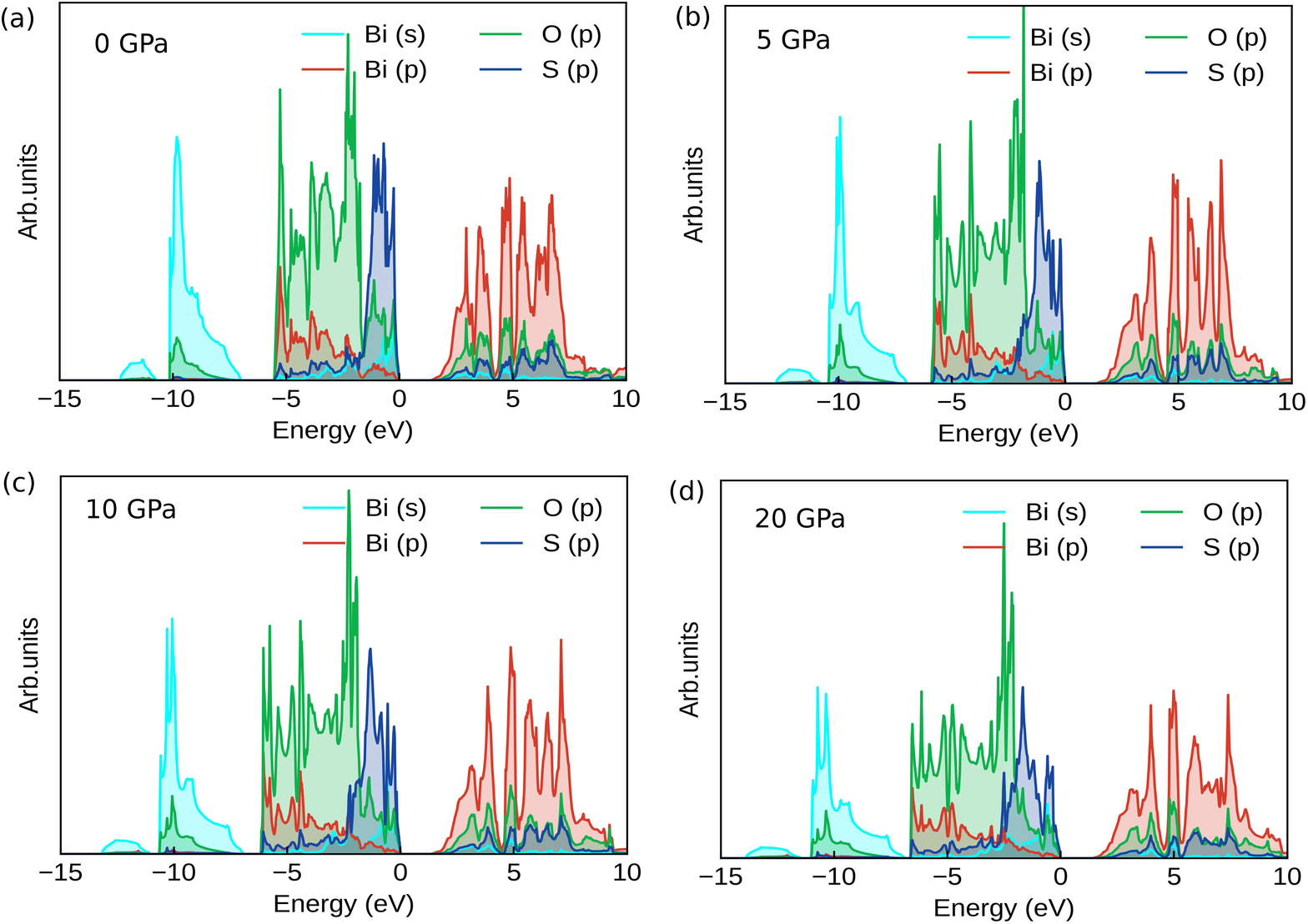}
   \caption{Calculated projected density of states at 0 GPa for ambient ($Pnmn$) phase and at 5 GPa, 10 GPa and 20 GPa for high pressure ($I4/mmm$) phase of Bi$_2$O$_2$S.}
    \label{fig:PDOS}
\end{figure*}{}

\begin{figure*}
    \centering
    \includegraphics[width=0.9\columnwidth]{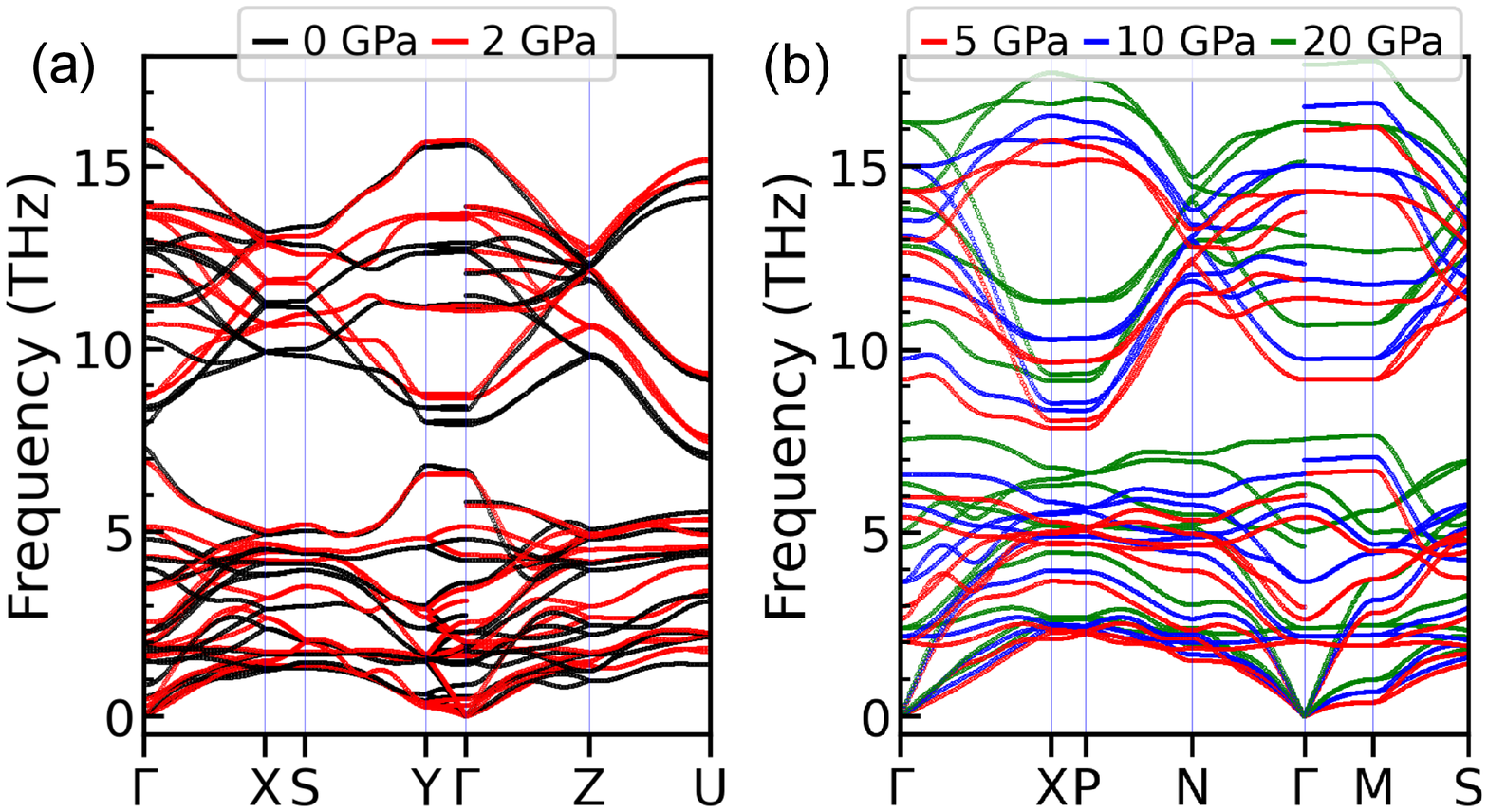}
   \caption{Calculated pressure dependent phonon dispersion curves of Bi$_2$O$_2$S in (a) $Pnmn$ (0, and 2 GPa), and (b) $I4/mmm$ (5, 10, and 20 GPa) phase.}
    \label{fig:phonon_SI}
\end{figure*}{}

\begin{figure*}
    \centering
    \includegraphics[width=1.0\columnwidth]{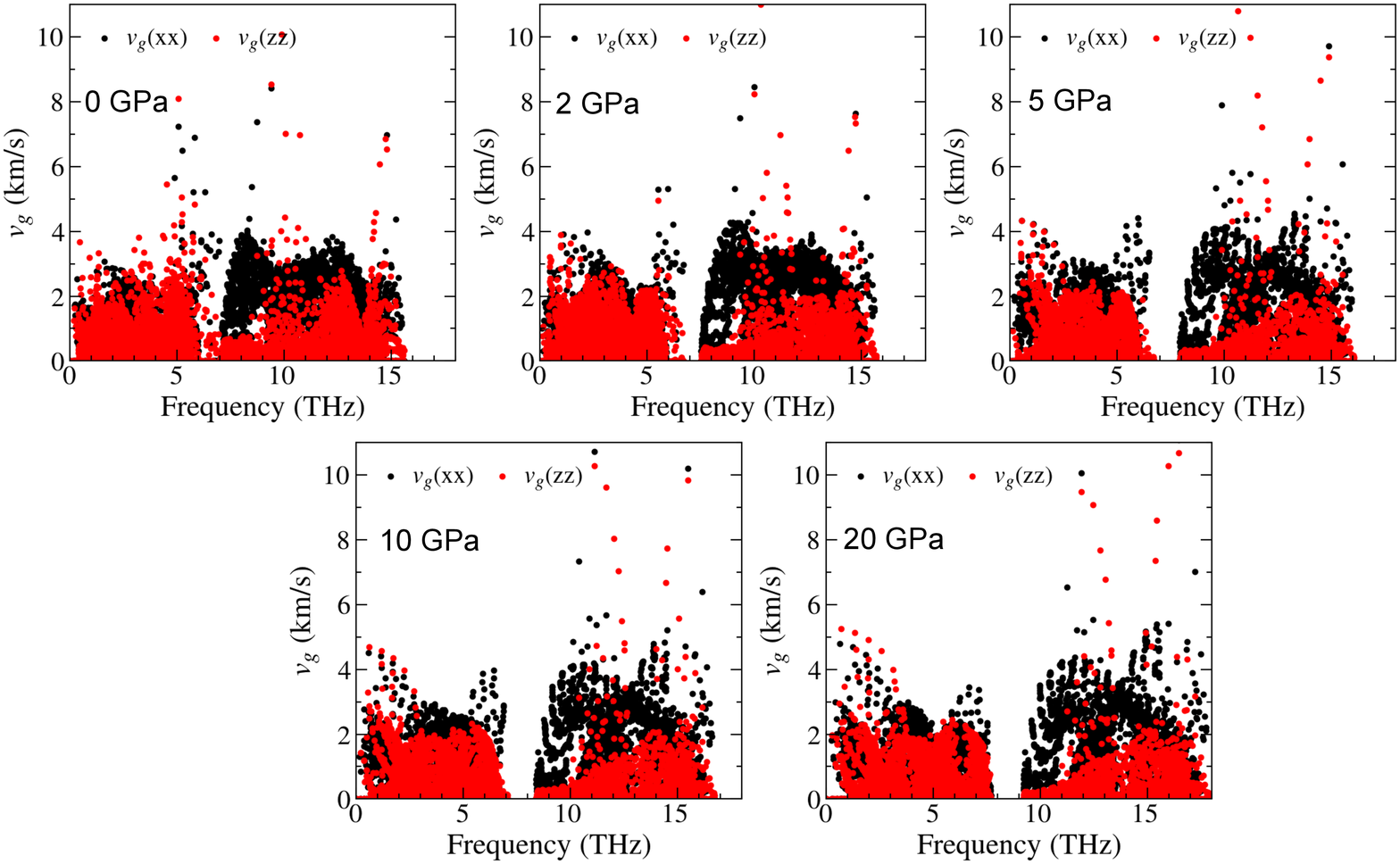}
   \caption{In plane (xx) and cross plane (zz) components of phonon group velocities as a function of frequencies at indicated pressures.}
    \label{fig:BL}
\end{figure*}{}

\begin{figure*}
    \centering
    \includegraphics[width=0.65\columnwidth]{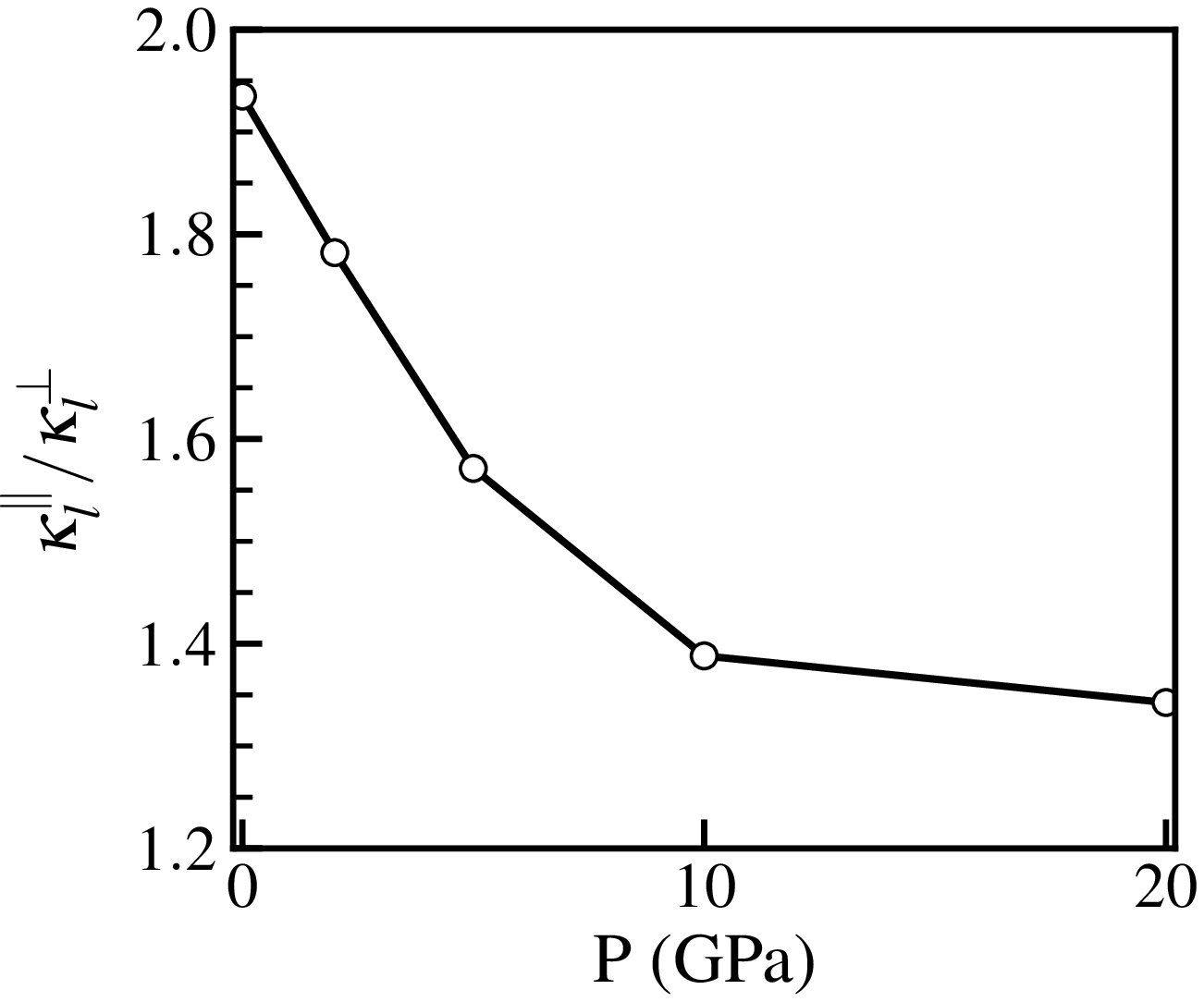}
   \caption{Effect of pressure on room temperature thermal transport anisotropy ($\kappa_{l}^{\parallel}/\kappa_{l}^{\perp}$)  as a function of pressure.}
    \label{fig:BL}
\end{figure*}{}
\clearpage
\begin{table*}[tbp]
\caption{Bond parameters such as  bond lengths (in $\AA$) and bond angles (in $^o$) of ambient ($Pnmn$) and high pressure ($I4/mmm$) phases of Bi$_2$O$_2$S obtained using DFT-D3 method.}
\label{table2}
\begin{tabular}{ccccc} \hline
Phase (Pressure) & Parameter    &    This work    &  Others$^a$ \\ \hline
$Pnmn$
(0 GPa)
&Bi-S1     &    3.00      &  3.149    \\   
&Bi-S2     &  3.51       & 3.227 \\  
&Bi-O1     &   2.24    & 2.288     \\
&Bi-O2     &   2.33    & 2.302       \\
&Bi-O3     &   2.43    & 2.316    \\
& $\angle$ S1-Bi-S1  &   80.00     &   \\
&$\angle$ S2-Bi-S2  &   66.64      &     \\
&$\angle$ S1-Bi-S2 ($\times$ 2)  &   73.73     &       \\
&          &   115.22    &   \\
&$\angle$ O1-Bi-O2  &    74.13    &       \\
&$\angle$ O2-Bi-O3  &    70.69    &       \\
&$\angle$ O2-Bi-O2  &    111.87    &       \\
&$\angle$ O1-Bi-O3  &    114.59    &         \\

$I4/mmm$
(5.6 GPa)
&Bi-S       &  3.14     &     \\  
&Bi-O        & 2.29      &     \\
&$\angle$ S-Bi-S ($\times$ 2)      &  74.21     &    \\      
&            &  117.12    &    \\ 
&$\angle$  O-Bi-O ($\times$ 2)      &  71.49     &     \\  
&             &  111.41     &    \\  \hline
\end{tabular}
\\ $^a$Ref. \cite{Hu2020}
\end{table*}

\clearpage
\bibliography{Refs.bib}